\documentclass[preprint2, natbib209]{aastex}

\usepackage{graphicx}
\usepackage{lscape}
\usepackage{rotating}

\shorttitle{Gas in Disks Around Sun-like Stars}
\shortauthors{Pascucci et al.}

\newcommand{\bd}{\begin{displaymath}} 
\newcommand{\ed}{\end{displaymath}} 
\newcommand{\mh}{H$_2$}  
\newcommand{\feii}{[Fe~{\sc ii}]}

\newcommand{\sii}{[S~{\sc ii}]}
\newcommand{\si}{[S~{\sc i}]}
\newcommand{\SiII}{[Si~{\sc ii}]}

\begin{document}
\bibliographystyle{apj}
\title{Formation and Evolution of Planetary Systems: Upper Limits to the Gas Mass in Disks Around Sun-like Stars}


\author{I. Pascucci\altaffilmark{1}, U. Gorti\altaffilmark{2}, D. Hollenbach\altaffilmark{3},
J. Najita\altaffilmark{4}, M.~R. Meyer\altaffilmark{1}, J.~M. Carpenter\altaffilmark{5},
L.~A. Hillenbrand\altaffilmark{5}, G.~J. Herczeg\altaffilmark{5}, D.~L. Padgett\altaffilmark{5},
E.~E. Mamajek\altaffilmark{6}, M.~D. Silverstone\altaffilmark{1}, W.~M. Schlingman\altaffilmark{1}, 
J.~S. Kim\altaffilmark{1}, E.~B. Stobie\altaffilmark{1}, 
J. Bouwman\altaffilmark{7}, S. Wolf\altaffilmark{7}, J. Rodmann\altaffilmark{7},
D.~C. Hines\altaffilmark{8}, J. Lunine\altaffilmark{9}, R. Malhotra\altaffilmark{9}}
\altaffiltext{1}{Steward Observatory, The University of Arizona, Tucson, AZ 85721.}
\altaffiltext{2}{University of California, Berkeley, CA 94720.}
\altaffiltext{3}{NASA Ames Research Center, Moffett Field, CA 94035.}
\altaffiltext{4}{National Optical Astronomy Observatory, Tucson, AZ 85719.}
\altaffiltext{5}{California Institute of Technology, Pasadena, CA 91125.}
\altaffiltext{6}{Harvard-Smithsonian Center for Astrophysics, Cambridge, MA 02138}
\altaffiltext{7}{Max Planck Institute for Astronomy, Heidelberg, Germany.}
\altaffiltext{8}{Space Science Institute, Boulder, CO 80301.}
\altaffiltext{9}{Lunar Planetary Laboratory, The University of Arizona, Tucson, AZ 85721.}







\begin{abstract}
We have carried out a sensitive search for gas emission lines at infrared
and millimeter wavelengths for a sample of 15 young sun--like 
stars selected from our dust disk survey with the {\it
Spitzer Space Telescope}. We have used mid--infrared lines
to trace the warm (300--100\,K) gas in the inner disk and millimeter 
transitions of $^{12}$CO to probe the cold ($\sim$20\,K)  outer disk. 
We report no gas line detections from our sample. Line flux upper limits are first
converted to warm and cold gas mass limits using simple
approximations allowing a direct comparison with values from the
literature. We also present results from more sophisticated models
following Gorti and Hollenbach (2004) which confirm and extend our
simple analysis. These models show that the \si{} line at 25.23\,\micron{} 
can set constraining limits on the gas surface density at the disk 
inner radius and traces disk regions up to a few AU.
We find that none of the 15 systems have more than 0.04\,M$_{\rm J}$ of gas 
within a few AU from the disk inner radius for disk radii from 1\,AU up to $\sim$40\,AU.  
These gas mass upper limits even in  the 8 systems younger 
than $\sim$30\,Myr suggest that most of the gas is dispersed early.
The  gas mass upper limits in the 10--40\,AU region, that is mainly traced by our CO data,
are $<$\,2\,M$_\oplus$.
If these systems are analogs of the Solar System,  either they have already
formed Uranus-- and Neptune--like planets or they will not form them beyond 100\,Myr.
Finally, the gas surface density upper limits at 1\,AU are smaller than 0.01\% of
the minimum mass solar nebula for most of the sources.  If terrestrial planets form frequently
and their orbits are circularized by gas,  then circularization occurs early. 

\end{abstract}


\section{Introduction}

Circumstellar disks are a natural outcome of the star formation process 
\citep{1987ARA&A..25...23S}.
Initially massive and gas-dominated, they evolve into tenuous dusty disks possibly
with embedded planets (e.g. \citealt{2006prpl.conf..Meyer}). 
Studying the evolution of gas and dust 
in circumstellar disks is essential to understanding how planets form.

The properties of circumstellar dust have been extensively studied in
young and old disks since the IRAS mission. The emerging evolutionary
picture is of sub-micron interstellar grains that grow to larger sizes 
(up to km-size bodies) and settle to the disk midplane  (e.g.
\citealt{2001A&A...375..950B,2003A&A...412L..43P,2005Sci...310..834A}).  
In a later stage, collisions between forming planets and/or minor bodies 
such as asteroids or Kuiper Belt Objects can produce second-generation dust
(or debris) in circumstellar disks 
(e.g. \citealt{2005ApJ...620.1010R,2005ApJ...632..659K,2006ApJ...636.1098B}).
This evolution leads to tenuous dust disks where the emission from
grains is optically thin to its own radiation. 
In contrast, less is known observationally about gas evolution 
in circumstellar disks.

The presence or absence of gas affects planet formation in profound ways.  Gas
masses between 1 and 10 times the dust masses in circumstellar disks control the
dust dynamics and shape the disk structure  (e.g. \citealt{2001ApJ...557..990T}).
The gas lifetime constrains the time for forming gas giant planets like Jupiter and Saturn.
In addition, the ultimate size and
distribution of giant planets may be affected by the disk gas mass. Even during the
assembly of terrestrial planets, which concluded approximately  30\,Myr
after the origin of the Solar System \citep{2005AREPS..33..531J}, a  few
Earth masses of gas in the terrestrial planet region could have influenced the eccentricity and final size of growing planets \citep{2004Icar..167..231K}. 

Observations of gas have mainly targeted young accreting disks (few Myr old) 
and probed the cold outer regions (outside $\sim$30\,AU) and the 
warm regions within $\sim$1\,AU of such disks.
Cold gas (20\,K$<T<$50\,K) is traced in the millimeter by CO rotational lines that 
indirectly measure gas masses in young accreting disks with radii of few hundred AU (e.g. \citealt{2004Ap&SS.292..407D}). 
The only bona--fide debris disk detected in millimeter CO transitions
is that around the A star 49~Cet \citep{1995Natur.373..494Z,2005MNRAS.359..663D}.
Warm gas ($\sim 1000$\,K) in the inner disk ($\leq$\,1\,AU) is probed in 
the near--infrared by CO vibrational transitions  
(e.g. \citealt{2003ApJ...589..931N}). 
These transitions may also trace  gas in  the terrestrial planet zone of 
evolved tenuous disks. In addition, H$_2$ rovibrational lines have been detected in a few
disks and found to trace the surface layer of the disks at distances of 10--30\,AU from the star 
(e.g. \citealt{2003ApJ...586.1136B}).
Observations in the ultraviolet have successfully detected a number of gas lines towards pre-main sequence stars 
(e.g. \citealt{2000ApJS..129..399V,2003ApJS..147..305V}). 
In particular fluorescent H$_2$ emission has been found tracing both warm 
gas ($\sim$\,2500\,K) at or near the disk surface within $\sim$1\,AU from the star \citep{2002ApJ...572..310H} as well as
surrounding molecular gas shocked in the interaction with stellar outflows (e.g. \citealt{2003AJ....126.3076W}). 
 These UV diagnostics are not sensitive to the bulk of the much colder gas at larger radii. This cold gas can be traced with H$_2$ absorption lines in the FUV but only in 
optically thin disks that are observed close to edge-on.  Stringent upper limits to the line-of-sight H$_2$ column density through the edge-on disks of $\beta$ Pic and AU Mic suggest that the primordial gas dispersed in less than 12\, Myr in both systems \citep{2001Natur.412..706L,2005ApJ...626L.105R}.

One region not adequately probed by most of the observations described above is the $\sim 1-30$ AU
region ($\sim 300-50$ K), corresponding to the  giant planet forming
region in the solar system. The Infrared Space Observatory (ISO) provided a 
first glimpse. \cite{2001Natur.409...60T} reported pure rotational 
H$_2$ S(0) and S(1) line detections from a large number of pre-main sequence stars and 
also from three main-sequence stars with debris disks. 
These detections translated into large reservoirs of gas, suggesting a gas dispersal timescale longer than the accretion timescale. However, subsequent 
ground-based infrared spectroscopy  \citep{2002ApJ...572L.161R, 2003MNRAS.343L..65S,2005ApJ...620..347S} and UV observations \citep{2001Natur.412..706L} cast doubt on whether the observed lines originated in disks. 

As part of the Formation and Evolution of Planetary Systems (FEPS)  {\it Spitzer} legacy program,
we are carrying out a comprehensive survey of disks around Sun-like stars to  characterize the
gas dissipation timescale. In conjunction with this survey, Gorti \& Hollenbach (2004, hereafter
GH04)  constructed detailed gas and dust models of optically thin dust disks in order to compare
observational results with model spectra.  In Hollenbach et al. 2005 (hereafter H05) we applied
these models to FEPS observations of the disk around the 30\,Myr star HD~105, and showed that
less than 1\,$M_{\rm J}$ of gas exists in the planet--forming region (between 1--40\,AU for a disk inner radius of $\sim$\,1\,AU). In the following,
we extend our analysis to 15 sun-like stars surrounded by optically thin dust disks,   most of
which have ages in the range 5--100\,Myr. The target selection and observational strategy are
presented in  Sect.~\ref{sect:obs}. From the non-detection of atomic and molecular gas lines,  we
place stringent upper limits on the gas mass in the planet--forming zone  (Sect.~\ref{sect:immres}
and \ref{mod}).   We discuss the implications on the gas evolution 
timescale and on the formation of gas giant and terrestrial planets in Sect.~\ref{sect:discuss}. 
Our findings are summarized in Sect.~\ref{sect:summ}.

\section{Observations and Data Reduction}\label{sect:obs}
In this Section we describe the target selection 
for our gas survey with {\it Spitzer} (Sect.~\ref{sect:tgselect}) 
and the data reduction of the
mid-infrared spectra (Sect.~\ref{sect:spitzer_redu}). In order to trace outer
disk regions with gas colder than 50\,K we complemented the {\it Spitzer}
data of many sources with millimeter observations of $^{12}$CO lines
(Sect.~\ref{sect:mmobs}).

\subsection{Target Selection and Observational Strategy}\label{sect:tgselect}  
We selected objects from the FEPS dusty disk survey of 328 stars based on two observational criteria and several ancillary parameters.  We chose the
nearest objects with ages mostly between 5 and 100\,Myr and  those 
located in the lowest infrared backgrounds so that
the observing time required to meet our goal would
be minimized.  Targets were chosen to span a range of infrared excess
emission, X--ray luminosity, and spectral type (within the bounds
of our program: 0.8--1.2 M$_{\odot}$).  We required
that mid-- to far--IR excess be detected in IRAS and ISO observations and extended the sample to excess sources discovered within FEPS.  We also included additional young sources 
($\leq\,30$\,Myr) lacking infrared excess emission. 
Here we focus on a subset of these sources, whose dusty disks are thin at optical wavelengths.
Results on the optically thick dust disks will be reported in a future contribution.
The properties of the optically thin sample are summarized in Table~\ref{stars}. 
Ten out of fifteen sources have excess emission from circumstellar dust starting at wavelengths longer than 
$\sim 20$\,\micron{}. Among the five sources with no excess emission, ScoPMS~214, AO~Men, and V343~Nor were included in the sample because of their young ages 
($\leq15$\,Myr). 
Our MIPS and low-resolution IRS observations do not confirm the excess emission reported in the literature towards the older two sources HD~134319 and HD~216803
\citep{2001ApJ...555..932S,1999ApJ...520..215F}.
In the case of HD~134319, the ISO 60$\mu$m flux was contaminated by a nearby source  detected in our MIPS 24 and 70\,$\mu$m exposures.

Our {\it Spitzer} IRS observations were designed to be sensitive to less than 
$\sim$0.5\,M$_{\rm J}$  in molecular hydrogen gas at a temperature of 100\,K. Integration times were set to achieve a 5-$\sigma$ 
detection against the noise. 
We used the IRS high-resolution modules providing R$\approx$700 spectral resolution 
over 9.9-37.2\,$\mu$m.  Note that the
natural line--width of the gas is expected to be 10-100 times smaller than the 
$\sim$430\,km/s resolution of the IRS high resolution spectra.  
The large wavelength coverage of the IRS on {\it Spitzer} enables us to probe various atomic 
and molecular lines in addition to the molecular hydrogen transitions.
In fact, GH04 show that certain lines, such as \si{}  at 25.23\,\micron{} and \sii{} at 34.8\,\micron , are expected in many instances to be significantly stronger than the H$_2$ lines.

\begin{table*}[bt]
\begin{center}
\caption{Summary of the target stellar parameters. Targets are ordered by distance, from the farthest (ScoPMS~214)
to the closest (HD~216803).}
\begin{tabular}{ll ccc ccc cc c}
\tableline\tableline
\# & Source    & RA        & DEC       &  SpT & Age 
   & D & $T_{\rm eff}$ & Log($L_{\star})$ &	Log($L_{\rm x}$) \\
   &       &[J2000]    &[J2000]    &       & [Myr] 
   & [pc]  &           [K] & [L$_{\sun}$] & [erg/s] \\
\tableline
1  &ScoPMS~214$^a$&16:29:49&-21:52:11.9& K0$^{(1)}$   & 5$^b$ &145 &5318&0.26  &30.72$^{(11)}$  \\
2  & MML~17       &12:22:33&-53:33:49.0& G0$^{(2)}$   &17$^c$ &124 &6000&0.43  &30.30$^{(2)}$   \\
3  & MML~28       &13:01:51&-53:04:58.1& K2$^{(2)}$   &17$^c$ &108 &4997&-0.35 &30.00$^{(2)}$      \\
4  & HD~37484     &05:37:40&-28:37:34.7& F3$^{(3)}$   &72    &60   &6727& 0.55 &29.07$^{(12)}$    \\
5  & HD~202917    &21:20:50&-53:02:03.1& G5$^{(4)}$   &30$^d$ &46  &5553&-0.18 &29.49$^{(13)}$    \\
6  &HD~134319$^a$ &15:05:50&+64:02:50.0& G5$^{(5)}$  &150    &44  &5716&-0.14 &29.23$^{(11)}$      \\
7  &HD~12039      &01:57:49&-21:54:05.3& G3/5$^{(6)}$ &30$^d$ &42  &5688&-0.05 &29.10$^{(13)}$   \\
8  &V343~Nor$^a$  &15:38:58&-57:42:27.3& K0$^{(4)}$   &12$^e$ &40  &5103&-0.03 &30.73$^{(11)}$  \\
9  &HD~377        &00:08:26&+06:37:00.5& G2$^{(7)}$   &88     &40  &5852& 0.09 &29.10$^{(11)}$   \\
10 & AO~Men$^a$   &06:18:28&-72:02:41.6& K3$^{(8)}$   &12$^e$ &38  &4359&-0.59 &30.16$^{(14)}$   \\  
11 &HD~209253     &22:02:33&-32:08:01.6& F6/7$^{(3)}$ &280    &30  &6217&0.21  &29.70$^{(15)}$    \\
12 &HD~35850      &05:27:05&-11:54:03.4& F7/8$^{(9)}$ &12$^e$ &27  &6138& 0.25 &30.60$^{(13)}$    \\
13 &HD~25457      &04:02:37&-00:16:08.2& F7$^{(10)}$   &110    &19  &6319& 0.32 &29.74$^{(16)}$  \\
14 &HD~17925      &02:52:32&-12:46:11.2& K1$^{(6)}$   &150    &10  &5173&-0.43 &28.97$^{(17)}$  \\
15 &HD~216803$^a$ &22:56:24&-31:33:56.1& K4$^{(3)}$   &400    &7.6 &4531&-0.71 &28.34$^{(17)}$    \\
\tableline
\end{tabular}\label{stars}
\tablecomments{Units of right ascension are hours, minutes, and seconds, and units of declination 
are degrees, arcminutes, and arcseconds. 
Spectral types are  from optical spectroscopy with accuracy of 0.5 to 1 subtype (references in parenthesis).
Column six reports the stellar ages: for members of stellar groups we report the mean group ages (see notes below), 
otherwise we report mean ages from Hillenbrand et al. (2006, in preparation). Typical errors are 30\% of the adopted 
age.
All distances, except for ScoPMS~214, are from Hipparcos
\citep{1997A&A...323L..49P}; we assume the mean distance to Upper~Sco 
 \citep{1999AJ....117..354D} for ScoPMS~214. 
  Effective temperatures are mainly from B,V,K photometry while bolometric luminosities are from 
 the best fit Kurucz stellar models to optical and near-infrared observations of the stellar photosphere 
 (we refer to Carpenter et al. 2006, in preparation for the procedure).
X-ray luminosities are reported in the last column of the table with references in parenthesis. }
\tablenotetext{a}{ These sources do not have excess emission at any of the observed wavelengths
(see Sect.~\ref{sect:tgselect} for details)}
\tablenotetext{b}{ This source belongs to the Upper~Scorpius group \citep{2002AJ....124..404P}}
\tablenotetext{c}{ These sources belong to the Lower Centaurus Crux association
\citep{2002AJ....124.1670M}}
\tablenotetext{d}{ These sources belong to the Tuc-Hor association
\citep{2000ApJ...535..959Z,2003ApJ...599..342S,2004ARA&A..42..685Z,2006astro.ph..3729M}}
\tablenotetext{e}{ These sources belong to the $\beta$~Pic moving group 
\citep{2001ApJ...562L..87Z,2004ARA&A..42..685Z}}
\tablerefs{ $^{(1)}$ \citet{1994AJ....107..692W}; $^{(2)}$ \citet{2002AJ....124.1670M};
$^{(3)}$ \citet{Houk_cat82}; $^{(4)}$ \citet{Houk_cat75}; 
$^{(5)}$ \citet{1998yCat.3206....0B}; $^{(6)}$ \citet{Houk_cat88}; 
$^{(7)}$ \citet{1978BICDS..15..121J}; $^{(8)}$ \citet{2006astro.ph..3770G};
$^{(9)}$ \citet{Houk_cat99}; $^{(10)}$ \citet{2003AJ....126.2048G}; 
$^{(11)}$ \citet{1999A&A...349..389V}; $^{(12)}$ \citet{1997A&A...326..221M};
$^{(13)}$ \citet{2003A&A...399..983W}; $^{(14)}$ \citet{2003AJ....126.1996M};
$^{(15)}$ \citet{1994A&A...285..272T}; $^{(16)}$ \citet{1998A&AS..132..155H}; 
$^{(17)}$ \citet{2004A&A...417..651S}; 
  }
\end{center}
\end{table*}

\begin{table*}[bt]
\caption{Log of the IRS high-resolution observations. 
 The $RAMP$ duration is the time in seconds per exposure. We acquired
$ncycles$ exposures before moving to the second slit position.
The total on--source integration time is $RAMP \times ncycles \times 2$.}
\begin{tabular}{l c cc}
\hline
\hline
\noalign{\smallskip}
\#    & AOR Key &    SH &  LH \\                  
      &         &  $RAMP \times ncycles$ & $RAMP \times ncycles$        \\
\hline
\hline
\noalign{\smallskip}
ScoPMS~214& 9776897 & 121.9$\times$6 & 60.95$\times$8	\\
MML~17    & 13463296& 121.9$\times$6 & 60.95$\times$6	  \\
MML~28    & 13462016& 121.9$\times$6 & 60.95$\times$7	\\
AO~Men    & 5458688 & 121.9$\times$2  & 60.95$\times$2    \\
HD~35850  & 9777920 & 31.46$\times$6 & 14.68$\times$10  \\
V343~Nor  & 5458944 & 121.9$\times$2  & 60.95$\times$2       \\
HD~12039  & 13461760& 121.9$\times$4 &  60.95$\times$3    \\
HD~202917 & 9778176 & 31.46$\times$5 & 14.68$\times$8	\\
HD~25457  & 9779712 & 6.29$\times$4  &  6.29$\times$2	\\
HD~37484  & 9780224 & 31.46$\times$10& 14.68$\times$10   \\
HD~377    & 13462272& 121.9$\times$3 &  60.95$\times$3  \\
HD~17925  & 9780480 & 31.46$\times$5 & 14.68$\times$8	\\
HD~134319 & 9779968 & 31.46$\times$6 & 14.68$\times$10  \\
HD~209253 & 9779200  & 31.46$\times$7  &  14.68$\times$4     \\
HD~216803 & 9777664 & 31.46$\times$5 & 14.68$\times$8	\\
\noalign{\smallskip}
\hline
\end{tabular}
\label{obspar}
\end{table*}

\subsection{Data Reduction of the High-resolution IRS Spectra}\label{sect:spitzer_redu}
Observations of our 15 targets were obtained between September 2004 and August 2005.  After
our validation observations of HD~105 in December 2003, in which we obtained only single 
on--source exposures, we switched to an on--source/off--source observing strategy to better
subtract the background and improve source extraction.  The objects have been all observed in
the Fixed Cluster--Offsets mode with two nod positions on--source (located  at 1/3 and 2/3 of
the slit length) and two additional sky measurements acquired just after the on--source
exposures.  The IRS or PCRS Peak-up options were used to place and hold the targets in the
spectrograph slit with positional uncertainties always better than 1$''$ (1 sigma radial). We
used the Short-High module (SH), covering the spectral range 9.9-19.6\,\micron, and the
Long-High module (LH) covering the wavelength range between 18.7 and 37.2\,\micron . Both
modules are cross--dispersed echelle spectrographs with spectral resolution of approximately
700.  A summary of the observational log is given in Table~\ref{obspar}.  The plate scale of
the detector is 2.3$''$/pixel for the SH module and 4.5$''$/pixel for the LH module.
Emission  at radii larger than  $\sim$50\,AU could be spatially resolved  with the SH module
for targets closer than 20\,pc  (see Table~\ref{stars}). However as we will show in
Sect.~\ref{gasopthin}, mid--infrared lines only trace  warm gas located within few AU from
the disk inner radius. Such emission is spatially unresolved for all targeted disks.

Raw high-resolution IRS data were processed with the Spitzer Science Center (SSC) pipeline S12.02.   We start our data reduction from the {\it droop} products.  Corrections applied in the SSC
pipeline at this stage include: saturation flagging, dark subtraction, linearity
correction, cosmic ray rejection and integration ramp fitting. Further data reduction
is based on self-developed IDL routines in combination with the SMART reduction
package developed by the IRS Instrument Team at Cornell (Higdon et al. 2004).  The
first step  consists of creating background subtracted images from sky measurements
acquired after the on--source exposures\footnote{Because we have the same RAMP and
number  of cycles for the source and the sky, we subtracted a sky exposure from each
on--source exposure}. Next we fix pixels marked bad in the "bmask" files with flag
value equal to $2^{9}$ or larger, thus including anomalous pixels due to
cosmic-ray saturation early in the integration, or preflagged as
unresponsive. The SSC provides additional hot pixel masks for individual
campaigns/modules that include permanently as well as temporarily hot pixels.
We correct for these hot pixels using the  irsclean
package provided by the SSC. The routines recognize if a pixel is marked in both the
bmasks and the  additional hot pixel masks and correct only once.   Bad and hot pixels
are cleaned by averaging spatial profiles in rows above and below  the affected row
and then normalizing the average profile to that of the affected row. The profiles are
fitted to the good data in the affected row by minimizing the $\chi^2$.  These
procedures have been extensively tested by the IRS GTO team and 
SSC\footnote{http://ssc.spitzer.caltech.edu/irs/roguepixels/}.
 We  use the background subtracted pixel--corrected images to extract 1D spectra
with the full aperture extraction routine in SMART.

\begin{figure*}
 \resizebox{\textwidth}{!}{\includegraphics[angle=0]{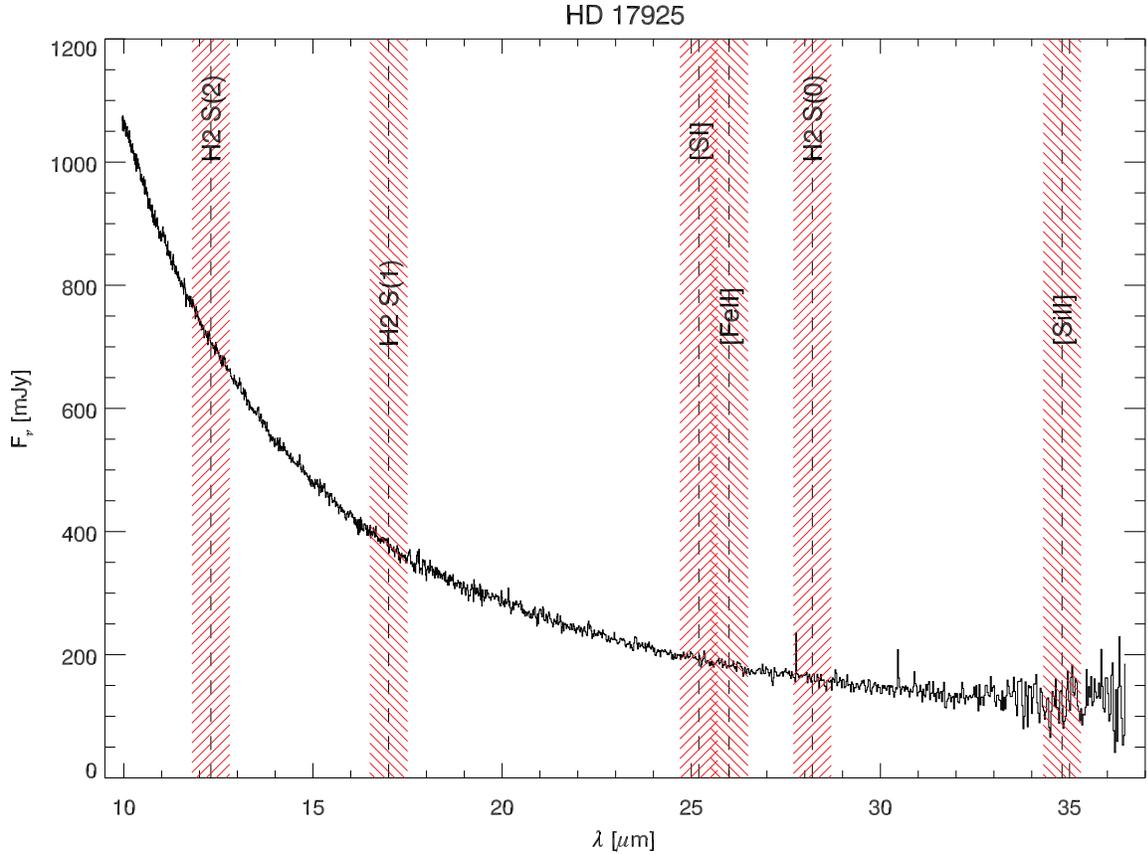}}
\caption{Spectrum of HD~17925 illustrating the location of the major molecular and atomic gas
transitions (black dashed lines) covered by the IRS high-resolution modules. 
The  hatched area (red in the electronic version) shows the region where we calculate
the line flux upper limits (see Sect.~\ref{upIRS} for more details).
The continuum between 10 and 40\,\micron{} is photospheric emission from HD~17925. Excess emission from dust was detected with ISO at 60\,\micron{} (\citealt{2001A&A...365..545H} and Fig.~\ref{figexcess2}).
The reduced spectra of all targets discussed  here are available in the on-line material.}
\label{hd_example}
\end{figure*}

We applied the same procedure to our sources and to
calibrators observed within the {\it Spitzer} IRS Calibration
program (PI, L. Armus). We used all the calibrators processed
with the S12.0.2 pipeline (2 observations of HD~166780, 3 of
HD~173511 and 1 of $\xi$~Dra)  to create two one-dimensional
spectral response functions (one at each nod position) from
the known stellar model atmosphere. The stellar models for the
calibrators are available at the SSC web
page\footnote{http://ssc.spitzer.caltech.edu/irs/calib/templ/
}. The extracted spectra of each source are then divided by
the spectral response function for each order and nod
observation. The lower right corner of the chip is not well
illuminated, resulting in  a drop of signal at the end of each
order. However, since the orders have a good overlap in
wavelength, we simply trimmed the low-signal regions,
retaining about 0.06\,\micron{} overlap in the SH module and 0.1 micron in
the LH module.  After trimming, spectra over all slit
positions and cycles are averaged on an order basis. The
uncertainties at each wavelength are estimated by the 1~sigma
standard deviations of the distribution  of the data points
used to calculate the mean spectrum over all cycles and nod
positions,  and thus indicate the repeatability of our
measurements.  To these uncertainties we included the error on the
derived spectral response function by taking the standard deviation of
the calibrator spectra divided by the stellar atmospheres.
The error quoted at overlapping wavelengths is
the error propagation of the mean value.  An example of a
reduced spectrum is shown in Fig.~\ref{hd_example}. All
reduced spectra are available in the on-line article.  We
report no detection of gas lines in any of these IRS spectra.
 We find that a few anomalous pixels remain in the reduced spectra that were not flagged
 in the bmasks nor in the hot pixel masks. These pixels are typically located
 at the beginning and/or at the end of a spectral order and not at the location of the expected gas lines
 (see for example the pixel at 27.78\,\micron{} in Fig.~\ref{hd_all_lines}).

\subsection{Millimeter CO Observations and Data Reduction}\label{sect:mmobs}
Observations of $^{12}$CO in J=2-1 (230.538~GHz) and J=3-2 (345.796 GHz) were obtained with the 10-m 
Submillimeter Telescope (SMT) during two campaigns, December-February 2003-2004 and
December-January 2005-2006. 
The full-width-half-maximum (FWHM) beam size of the SMT observations is 
33$''$ and 22$''$ for the J=2-1 and J=3-2 transitions respectively.
The data were recorded using the CHIRP digital spectrometer (bandwidth of $\sim$215\,MHz and
resolution of $\sim$40\,kHz) and 
reduced using CLASS in the GILDAS\footnote{http://www.iram.fr/IRAMFR/GILDAS}
data reduction package. 
During data reduction, the spectra were smoothed to a velocity resolution of 
0.2~km~s$^{-1}$, about ten times smaller than the expected width of the line. A linear baseline was then removed from each spectrum,
and all the spectra for a given source were coadded by weighting each spectra
by rms$^{-2}$. We calibrated our temperatures by measuring the ratio of observed and expected emission lines from planets (the latter was computed with the ASTRO software within GILDAS).
These ratios, also known as beam efficiencies, had mean values of 0.77 and 0.50 for the
J=2-1 and J=3-2 during the first campaign and 0.74 and 0.55 for the J=2-1 and J=3-2 during the second campaign.
Repeated observations of spectral line calibrator sources during each campaign indicate the 
dispersion in the calibration is 19\%, and 28\% for J=2-1 and J=3-2  respectively.
 This dispersion was not included in the estimates of the line flux upper limits in Sect.~\ref{upCO}.
For two FEPS sources (HD~37484 and
ScoPMS~214), the heliocentric velocity was unknown at the time of the 
observations, and the spectrometer was centered on 0.0~km/s. The
velocities were later determined from high resolution optical spectra 
\citep{2005...white}, and in all three instances, the
velocities were well within the spectrometer bandwidth ($>$100\,km/s).
We did not detect any CO emission line from any of our sources. 
The 1-sigma upper limits per channel are provided in Table~\ref{tab:co}.
Values are given in the main beam scale, 
i.e. the antenna temperatures have been divided by the main beam efficiencies of the telescope at the specific frequencies.


\begin{table*}[bt]
\begin{center}
\caption{Millimeter observations and estimated 1-sigma upper limits (per channel) in K.}
\begin{tabular}{l c| cc}
\hline
\hline
\noalign{\smallskip}
Source    & $v_{\star}^a$ & \multicolumn{2}{c}{$\Delta T_{\rm rms}$ [K]} \\                 
          &  [km/s]       & CO(2-1) &  CO(3-2)    \\
\tableline
\tableline
\noalign{\smallskip}
ScoPMS~214    & -7.6 & 0.033$^{I}$ & 0.18      \\
HD~35850      & 20.2 & 0.043 & 0.083$^{I}$  \\
HD~12039      & 3.1  & 0.049 & 0.078\\
HD~25457      & 17.6 & 0.028$^{I}$  & 0.044       \\
HD~37484      & 23.7 & 0.038$^{I}$  & --       \\
HD~377        & -4   & 0.043  & 0.13$^{I}$       \\
HD~17925      & 17.7 & 0.042$^{I}$  & 0.056       \\
HD~134319     & -6.8 &0.06$^b$& 0.068$^{I}$  \\
HD~209253     & 8.0  & --     & 0.095$^{I}$  \\
\noalign{\smallskip}
\tableline
\end{tabular}
\tablenotetext{a}{v$_{\star}$ are the radial heliocentric velocities of the stars as measured by stellar spectroscopy \citep{2005...white}. 
Errors on $v_{\star}$ are about 0.5\,km/s for all sources except for HD~377 where our uncertainty is 2\,km/s.}
\tablenotetext{b}{ HD~134319 has been observed in the CO(2-1) transition by
Najita \& Williams using the heterodyne receiver at JCMT telescope (priv. communication, for the data reduction we refer to \citealt{2005ApJ...635..625N}) }
\tablenotetext{I}{Results from the first observational campaign (December-February 2003-2004).}
\label{tab:co}
\end{center}
\end{table*}

\section{Simple Estimates of Gas Mass Upper Limits}\label{sect:immres}
We have not detected any lines associated with the gas in our IRS spectra, nor in our millimeter data. 
Line flux upper limits can be compared to model predictions of line strength and provide stringent 
constraints on the amount of gas in the planet--forming region. 
We pursue rigorous theoretical comparisons in Sect.~\ref{gasopthin}.
Here for comparison to previous studies, we estimate gas mass upper limits in the approximation of 
optically thin emission at assumed fixed temperatures. 
Upper limits to the warm and cold gas are derived from the H$_2$ (Sect.~\ref{upIRS}) and CO non-detections 
respectively (Sect.~\ref{upCO}). In addition, we calculate the upper limit on
the radius of a putative gas disk under the assumption that the CO emission
is optically thick in the inner regions (Sect.~\ref{upCO}). This radius will
be later compared to that derived from our detailed modeling of gas
disks which satisfy the observed infrared line flux limits.

\subsection{Warm Gas Limits from H$_2$ Observations}\label{upIRS}
\begin{figure*}
 \resizebox{\textwidth}{!}{\includegraphics[angle=90]{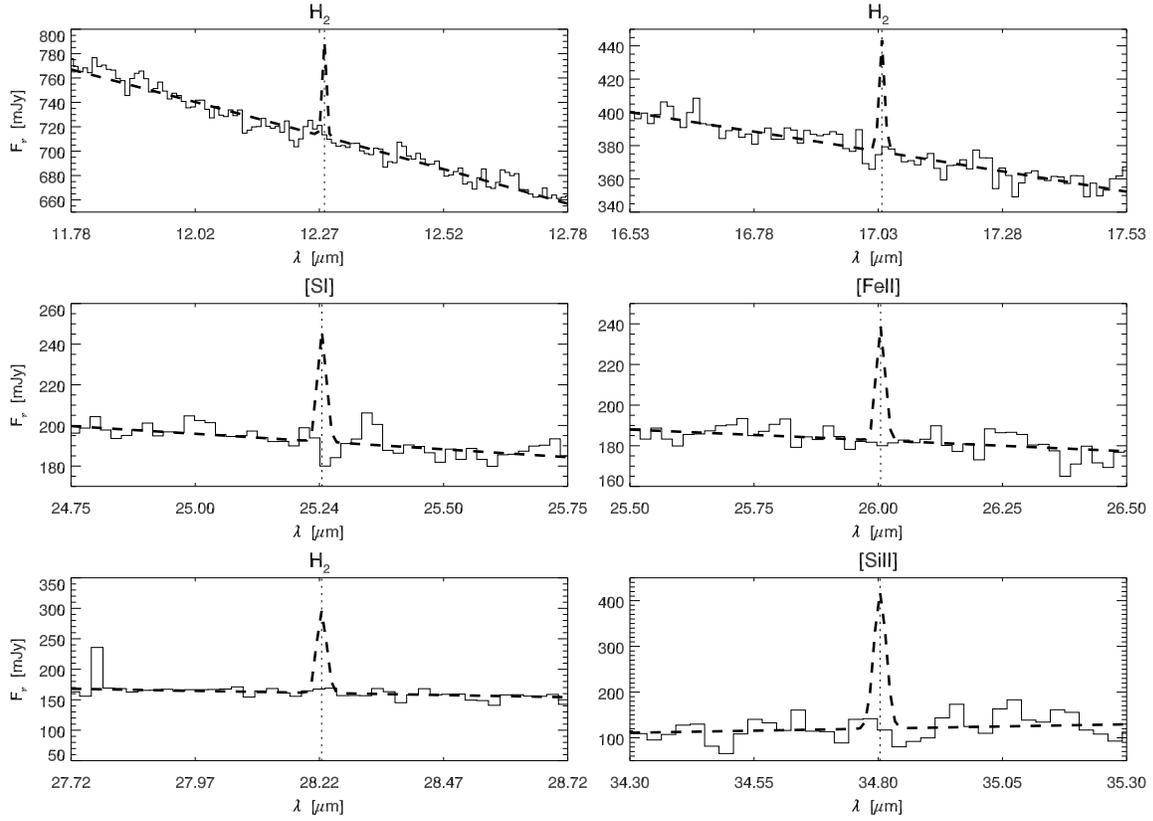}}
\caption{Expanded view of the wavelength regions around the expected lines toward the source
HD~17925. 
Dashed lines are the hypothetical 5\,$\sigma$ line fluxes calculated assuming a 
resolution of 700 at all wavelengths. Lines are expected to be unresolved at all
wavelengths. The pixel at 27.78\,\micron{} is an example of
anomalous pixel that was not flagged in the bmasks nor in the hot pixel masks.}
\label{hd_all_lines}
\end{figure*}

 To estimate line flux upper limits from the mid--infrared observations, we first
extract a $\pm$0.5\,\micron{} region around the expected feature. Then
we fit a baseline using a first order polynomial.
Five sigma upper limits to the line flux  are derived by taking
the local RMS dispersion of the pixels in the baseline subtracted spectrum over two 
pixels per resolution element, assuming the noise is uncorrelated.
Note that because we calibrate our spectra from the average spectral response
function of various standard stars, our upper limits partly include the error in the absolute flux
calibration.
We have tested  that the line flux  upper limits are not very sensitive to the 
width of the region where we fit the baseline. 
Choosing a region around the line that is 0.5\,\micron{} smaller or larger than the 1\,\micron{} 
width we adopt changes the upper limits on average by only 6\%\footnote{These tests have been made on the three lines that are the most informative  when compared to the gas models:  the  \mh~S(1)  at 17.04\,\micron , the
\si{} at 25.23\,$\mu$m, and the \feii{} at 26.00\,$\mu$m}. 
Our approach provides conservative upper limits. The use of a higher order
 polynomial to fit the continuum and/or the inclusion of the errors at each wavelength would
 lower the computed RMS and thus the line flux upper limits. Our RMS values are on average 20\% higher than the error at the wavelengths of the expected lines.
Table~\ref{line-fluxes} summarizes the line flux upper limits 
for many of the transitions observed in the IRS modules:
\mh~S(2)  at 12.28\,$\mu$m, \mh~S(1)  at 17.04\,$\mu$m, \mh~S(0) at 28.22\,$\mu$m,
\si{} at 25.23\,$\mu$m, \feii{} at 26.00\,$\mu$m,  and \SiII{} at 34.80\,$\mu$m. 
We show examples of hypothetical 5\,$\sigma$ lines in Fig.~\ref{hd_all_lines}.

\begin{table*}[bt]
\caption{Line flux upper limits (5-sigma) from the high-resolution IRS spectra}
\begin{tabular}{l cccccc}
\tableline\tableline
Source& \multicolumn{6}{c}{Log[Line flux\, (W/cm$^2$)]} \\
      &  \mh S(2) & \mh S(1) &  \mh S(0) & \si & \feii & \SiII \\
\tableline
ScoPMS~214&  -21.73 &-21.89 &-21.41 &-21.57 &-21.57  &-20.99\\
MML~17    &  -21.86 &-21.99 &-21.56 &-21.46 &-21.62  &-20.95\\
MML~28    &  -21.95 &-22.02 &-21.51 &-21.60 &-20.27  &-20.81\\
HD~37484  &  -21.13 &-21.28 &-21.20 &-21.18 &-21.36  &-20.76\\
HD~202917 &  -21.10 &-21.11 &-21.15 &-21.11 &-21.34  &-20.73\\
HD~134319 &  -21.18 &-21.27 &-21.17 &-21.15 &-21.28  &-20.73\\
HD~12039  &  -21.60 &-21.74 &-21.29 &-21.01 &-21.42  &-20.78\\
V343~Nor  &  -21.29 &-21.52 &-21.00 &-21.18 &-21.38  &-20.60\\
HD~377    &  -21.45 &-21.66 &-21.33 &-20.95 &-21.12  &-20.53\\
AO~Men    &  -21.60 &-21.74 &-21.25 &-21.19 &-21.17  &-20.87\\
HD~209253 &  -20.82 &-20.86 &-21.07 &-20.98 &-21.21  &-20.51\\
HD~35850  &  -20.94 &-21.07 &-21.07 &-20.97 &-21.18  &-20.75\\
HD~25457  &  -20.49 &-20.57 &-20.42 &-20.66 &-20.77  &-20.11\\
HD~17925  &  -20.80 &-21.01 &-20.89 &-21.24 &-21.24  &-20.64\\
HD~216803 &  -20.80 &-20.98 &-21.06 &-20.93 &-21.14  &-20.68\\
\tableline
\end{tabular}\label{line-fluxes}
\end{table*}

We convert H$_2$ line flux upper limits into upper limits on the H$_2$ mass (see e.g.
\citealt{2001ApJ...561.1074T} for the equations) assuming that the gas is in local 
thermal equilibrium, that the H$_2$ lines are optically thin, and  that the 
gas temperature is 100, 150 or 200\,K.  Temperatures of $\approx$100\,K correspond to the gas from which our own
Jupiter and Saturn are thought to have formed (e.g. \citealt{2001ApJ...550L.227G}).
 Expanding the temperature range to 200\,K  allows us to make a direct comparison
with published  gas masses and upper limits for circumstellar disks (e.g.
\citealt{2001ApJ...561.1074T,2005ApJ...620..347S,2005astro.ph.11657C}; H05; \citealt{2006astro.ph..5277C}) 
and illustrate how gas masses depend on the assumed gas temperature. 
 The \mh~S(2)
line does not set stringent  limits for temperatures lower than $\sim$300\,K, the
\mh~S(1) transition is the most sensitive line at 150 and 200\,K, while the \mh~S(0) 
and the \mh~S(1) transitions provide similar values for gas at 100\,K. 
In Table~\ref{umass} we report the lowest gas mass upper limits 
for temperatures of 100, 150, and 200\,K and indicate whether they are derived from 
the \mh~S(0) or the \mh~S(1) transitions for gas at 100\,K. 
 This simple approach shows that gas masses are very sensitive to the assumed gas temperature:
if the emitting gas is as cold as 100\,K it takes about hundred times more mass to produce the measured line flux upper limits than if gas is at 200\,K. 
Hence, it is very important to utilize detailed disk thermal models which describe the
gas density and temperature distribution and which predict line fluxes from
disks of a given mass for comparision with the observed upper limits on line fluxes.

\begin{table*}[bt]
\caption{Gas mass upper limits estimated from the \mh~S(0) and \mh~S(1) non--detections
for three gas temperatures. We used ortho--to--para ratios of 1.6, 2.5, and 3 for gas temperatures of 100,
150, and 200\,K respectively \citep{1999ApJ...516..371S}.} 
\begin{tabular}{l cc}
\tableline
\tableline
\noalign{\smallskip}
Source  & \mh{} mass  [$M_\oplus$]    & Most sensitive line$^a$ \\
        & for 100--150--200\,K gas        &  for 100\,K gas\\
\tableline
\noalign{\smallskip}
ScoPMS~214 &  200--6.6--1.3 &  \mh~S(1) \\
MML~17     &  115--3.8--0.8 &  \mh~S(1) \\
MML~28     &  82--2.7--0.6  &  \mh~S(1)\\
HD~37484   &  64--4.6--0.9  &  \mh~S(0) \\  
HD~202917  &  42--4.0--0.8  &  \mh~S(0) \\
HD~134319  &  37--2.5--0.5  &  \mh~S(0)   \\
HD~12039   &  24--0.8--0.2  &  \mh~S(1)  \\
V343~Nor   &  36--1.2--0.2  &  \mh~S(1)\\
HD~377     &  21--0.9--0.2  &  \mh~S(0) \\
AO~Men     &  19--0.6--0.1  &  \mh~S(1) \\
HD~209253  &  22--3.1--0.6  &  \mh~S(0) \\ 
HD~35850   &  17--1.5--0.3  &  \mh~S(0) \\
HD~25457   &  39--2.4--0.5  &  \mh~S(1) \\  
HD~17925   &  4--0.2--0.05  &  \mh~S(0) \\  
HD~216803  &  1--0.1--0.03  &  \mh~S(0)\\ 
\tableline
\tablenotetext{a}{For gas temperatures of 150 and 200\,K the \mh~S(1) line is always
more sensitive than the \mh~S(0) and the \mh~S(2) lines. }
\end{tabular}
\label{umass}
\end{table*}

Our measurements for optically thin disks are more sensitive than the ISO observations  by
\citet{2001ApJ...561.1074T} by at least a factor of 4 for sources at similar distance. The line
detections claimed by Thi et al. suggested that relatively large amounts  (6.7-0.2\,$M_{\rm J}$) of gas
can persist into the debris--disk phase.  Our gas mass upper limits show that even the six youngest disks
in our sample  (5-20\,Myr) have less than 0.6 Jupiter masses of 100\,K gas (or less than 0.005 Jupiter
masses of 200\,K gas). Our results are in agreement with the recent findings by
\citet{2005astro.ph.11657C} who reports masses of 100\,K gas smaller than 15$M_\oplus$ (or
0.05\,$M_{\rm J}$)
in the disks of $\beta$~Pictoris ($\sim$12\,Myr, \citealt{2001ApJ...562L..87Z}) and 49~Ceti 
($\sim$10\,Myr, \citealt{1995Natur.373..494Z}). 
These low gas masses argue for short gas dispersal timescales.

Gas masses in Table~\ref{umass} are provided  mainly for comparison to previous studies. 
More detailed modeling of the dust and gas components is necessary to  understand the
mechanisms relevant to the gas heating, to determine gas temperatures, and thus infer total gas masses from our non-detections. This procedure has been fully explored in H05 for HD~105, one of the first sources in  the FEPS gas program, and is extended to our larger sample in Sect.~\ref{mod}.

\subsection{Cold Gas Limits from CO Observations}\label{upCO}

For the millimeter observations, we convert the 1-sigma noise into line flux upper limits as follows:

\begin{equation}
\Delta F(\nu) =  \frac{2\, k \,\Delta T_{\rm rms}}{\lambda^2} \times \frac{\pi\,\theta^2}{4} \times  d\nu \times \sqrt N
\end{equation}

\noindent where $k$ is the Boltzmann constant, $\Delta T_{\rm rms}$ are our 1-sigma
upper limits per channel as  reported in Table~\ref{tab:co},  $\theta$ is the beam
in radians at the transition $\lambda$, $d\nu$ is the frequency resolution ($dv$ =
0.2 km/s),  and $N$ is the number of channels on which the line is distributed.
Assuming a line width of 5\,km/s, which is the Keplerian velocity for gas at
$\sim$\,30AU around a solar mass star, we have 25 channels for the SMT
resolution of 0.2\,km/s.  The 5-sigma line flux upper limits are summarized in
Table~\ref{tab:upCO} and will be used here to provide an upper limit either to the
CO mass in disks optically thin to the CO transitions or to the size of an optically
thick CO disk. We will use our gas models to fully characterize the CO emission in
Sect.~\ref{gasopthin}.

Assuming that CO is optically thin throughout the disk, we can convert our line flux
upper limits into CO mass upper limits. We use the formulation of 
\citet{1986ApJ...303..416S} with appropriate changes to take into account different
CO transitions and adopt an   excitation temperature of 20\,K. We note that the CO
mass is not very sensitive to the gas temperature and changes only by $\sim$20\%
if we assumed an excitation temperature of 40\,K.   Five-sigma upper limits to the CO mass are
summarized in Table~\ref{tab:upCO}.   Our limiting values are typically a factor of
10 higher than those reported by \citet{2005ApJ...635..625N} for the disks around
HD~104860 and HD~107146, due to a combination of shorter integration times and the
assumption of larger line widths.  Still these upper limits convert to less than
a few $M_{\oplus}$ of total gas mass for many sources assuming the interstellar
H$_2$/CO number ratio of 10$^{4}$.  However, there are at least two important issues
that need to be considered: the  possible condensation of CO onto grains and the
photodissociation of CO molecules. Both these processes reduce the CO gas phase
abundance relative to H and thereby raise the upper limit on the total mass.
Condensation of CO occurs for grain temperatures $\leq$\,50\,K depending on the
substrate onto which CO is adsorbed (see e.g. \citealt{2005ApJ...635..625N} for a
discussion). The lack of submillimeter continuum data for our sample does not allow  us to
constrain the temperature of the cold grains, and thereby to estimate the effect of CO
condensation. However, we will show in Sect.~\ref{mod} that photodissociation is a
severe problem for tenuous gas disks  such as those we  have observed. If CO is optically thin
but much of the gas phase carbon is found in C$^+$ or C, then the CO gas we detect
would be representative of only a small fraction of the  total gas mass, and the
total gas mass associated with the optically thin CO could be $>> $ few $M_{\oplus}$. 
In addition, the optically thin limits on CO mass tell us nothing about the 
CO mass located in the optically thick inner regions.

Assuming the existence of such an optically thick inner disk (which is indeed
present in young accreting disks,  \citealt{2004come.book...81D}), we can set limits
to the disk size by our CO flux upper limits. For simplicity we describe the CO
emission as that from a blackbody times the solid angle subtended by a 
 face--on disk\footnote{Because we do not have information on the disk inclinations, we assume face--on configurations here and in the following gas models. 
For optically thick lines the observed flux is proportional to the projected area of the disk. Thus,  fluxes from optically thick lines can be corrected by multiplication with the cosine of the disk inclination.}  
and compare this emission to our 5--sigma line flux upper limits. Table~\ref{tab:upCO}
summarizes the inferred upper limits on optically thick disk radii for a black body
temperature of  20\,K. Disk radii are more sensitive to the assumed blackbody temperature than
disk masses: a blackbody temperature of 40\,K results in $\sim$35\% smaller radii than those reported in 
Table~\ref{tab:upCO}. It is important to characterize the region inside which
optically thick CO gas could exist because inside that region, gas masses could be
high and still be  undetected in CO.  It is in these regions that the infrared
upper limits may constrain the mass or surface density.  In Section~\ref{mod} we
will compare these radii, calculated for constant T, with those inferred by detailed
gas models which compute the  radial and vertical dependence of T and the detailed
radiative transfer operating for the CO transitions as well as for the infrared transitions.

\begin{table*}[bt]
\begin{center}
\caption{Five-sigma upper limits for the line fluxes, CO masses, and outer disk radii. }
\begin{tabular}{l cc| cc | cc}
\hline
\hline
\noalign{\smallskip}
Source    &\multicolumn{2}{c|}{Log(Line flux)} & \multicolumn{2}{c}{$M_{\rm CO}$} 
&\multicolumn{2}{c}{$R_{\rm out}$}\\  
          &\multicolumn{2}{c|}{[W/cm$^2$]}     & \multicolumn{2}{c}{[M$_\oplus$]} 
          & \multicolumn{2}{c}{[AU]} \\
          &  CO(2-1) &  CO(3-2)        & CO(2-1) &  CO(3-2)  & CO(2-1) &  CO(3-2)   \\
\tableline
\noalign{\smallskip}
ScoPMS~214   & -23.43   & -22.51  & $1.2\times 10^{-2}$  & $3.1\times 10^{-2}$ &236 & 403\\
HD~37484     & -23.37   & --      & $2.4\times 10^{-3}$  & --                  &105 & --\\
HD~134319    & -23.17   & -22.93  & $2.0\times 10^{-3}$  & $1.1\times 10^{-3}$ & 97 & 75\\
HD~12039     & -23.26   & -22.87  & $1.5\times 10^{-3}$  & $1.1\times 10^{-3}$ & 83 & 77 \\
HD~377       & -23.32   & -22.65  & $1.2\times 10^{-3}$  & $1.7\times 10^{-3}$ & 74 & 94\\
HD~209253    &  --      & -22.79  & --    &   $6.9\times 10^{-4}$              & -- & 61\\
HD~35850     & -23.32   & -22.85  & $5.4\times 10^{-4}$  & $4.9\times 10^{-4}$ & 50 & 51\\
HD~25457     & -23.50   & -23.12  & $1.7\times 10^{-4}$  & $1.3\times 10^{-4}$ & 28 & 26\\
HD~17925     & -23.33   & -23.02  & $7.2\times 10^{-5}$  & $4.5\times 10^{-5}$ & 18 & 15\\
\noalign{\smallskip}
\tableline
\end{tabular}
\tablecomments{CO masses are calculated assuming optically thin emission. The outer disk radii are estimated from the 
assumption of CO being optically thick.  For both estimates we adopted a gas temperature of 20\,K.}
\label{tab:upCO}
\end{center}
\end{table*}

\section{Constraining the Gas Mass in the Planet--forming Zone}\label{mod}
Even though H$_2$ is  the dominant constituent of the gas, H$_2$ lines are
not always the strongest transitions from  circumstellar disks.  Depending
on the density, temperature and chemical structure (which
in turn depends on the radiation field and disk surface density
distribution), other infrared atomic and molecular lines  can have
higher luminosities (GH04). In this section, we make use of the line flux
upper limits derived from  the mid-infrared and mm CO transitions to derive
constraining upper limits for the gas mass in  the planet--forming zone.

\citet{2005ApJ...631.1180H} have shown that the dust disk of HD~105  is so tenuous that the dust
does not affect the gas temperature or gas disk structure. In
Sect.~\ref{s_thindust} we demonstrate that our systems also have too low
dust surface density for gas-grain collisions to heat or cool the gas. Gas
heating
 in these tenuous disks is dominated by X-rays and UV radiation
from the central star, and therefore
 we can ignore the dust component
in modeling the gas emission (Sect.~\ref{gasopthin}).  This is completely
opposite to the case for young optically thick disks, where the gas
temperature in much of the disk is coupled to the dust temperature by
collisions, and where the surface gas is heated by the grain photoelectric
heating mechanism (e.g. \citealt{2004A&A...428..511J}).

\begin{figure}
 \resizebox{\hsize}{!}{\includegraphics[angle=0]{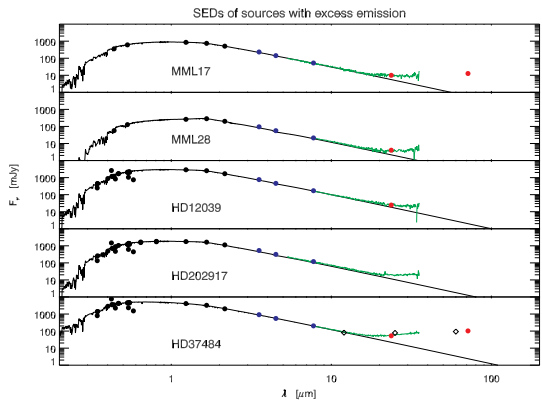}}
\caption{Spectral energy distributions (SEDs) of sources with excess emission.
Blue, red and black filled-circles are IRAC, MIPS and ground-based data points.
Green lines are IRS low resolution spectra. 
IRAS and ISO measurements are in diamonds and squares  respectively. 
Kurucz model atmospheres are overplotted with black lines. }
\label{figexcess1}
\end{figure}

\begin{figure}
 \resizebox{\hsize}{!}{\includegraphics[angle=0]{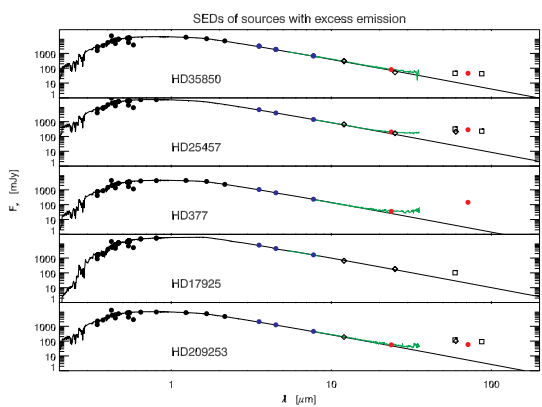}}
\caption{SEDs of sources with excess emission.
See Fig.~\ref{figexcess1} for the plotting symbols.}
\label{figexcess2}
\end{figure}

\subsection{Gas--dust Collisions versus X--ray Heating}\label{s_thindust}

\begin{figure*}
 \resizebox{\hsize}{!}{\includegraphics[angle=0]{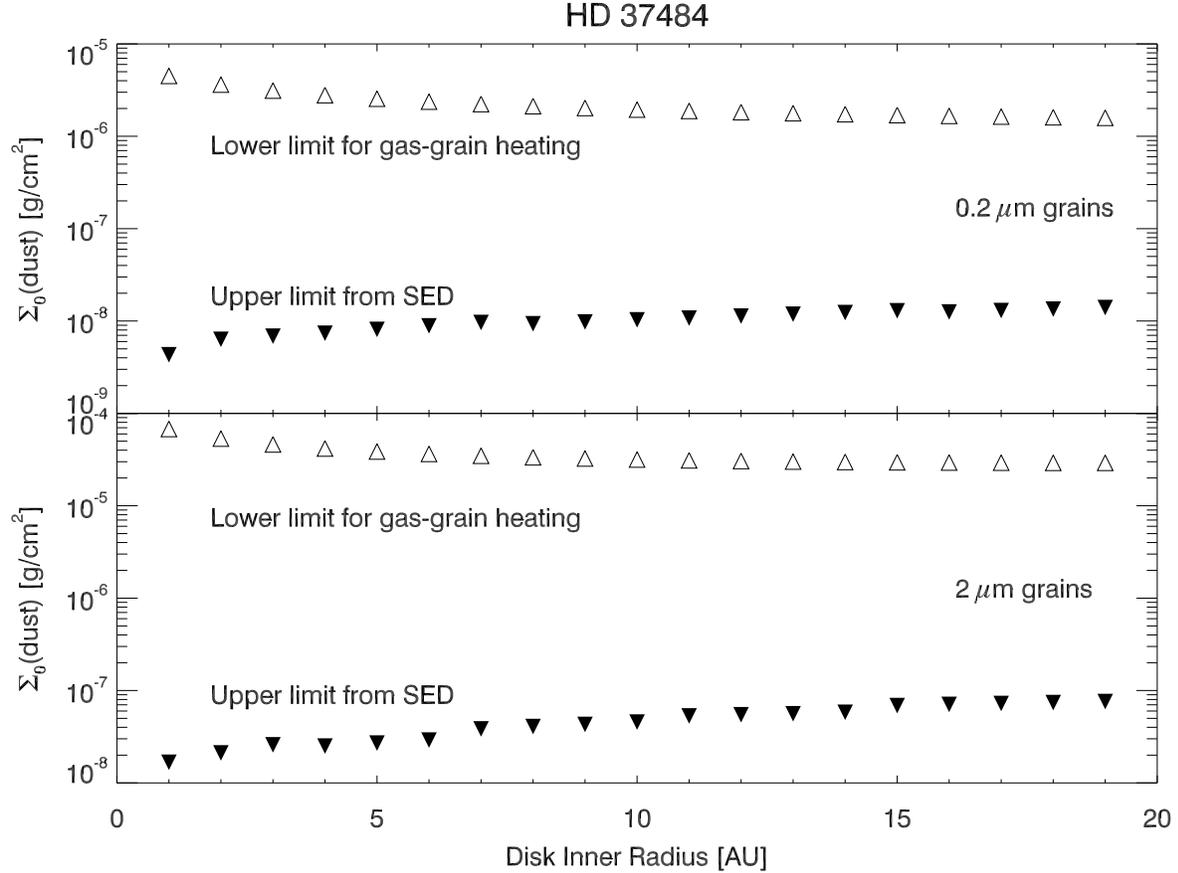}}
\caption{Comparison of dust surface densities at different disk inner radii
for the source HD~37484.
Open triangles  indicate the minimum surface density for  dust to contribute as much as 
X-rays in heating the gas at 70\,K. Upside down filled triangles
are the surface density upper limits imposed by the observed excess emission.
For the dust, we adopted silicate grains with two sizes (0.2 and 2\,\micron) and opacities from \citet{1984ApJ...285...89D}.
Note that there are at least  two orders of magnitude between the limits  inferred from the SED modeling and from 
gas--grain heating at any disk inner radius. This shows that gas-grain collisions are an insignificant heating source in these
tenuous dust disks. They also cannot dominate the cooling of the X-ray heated gas (see text).}
\label{figsurf}
\end{figure*}

To understand whether dust may be an important heating mechanism for the gas, we 
need to evaluate the dust surface density of our disks. 
The observed infrared continuum excess from a disk gives an upper limit to the 
dust surface density for a chosen grain size. 
Ten of our gas sources have infrared excess emission at wavelengths longer 
than $\sim$20\,\micron{} with spectral energy distributions (SEDs) characteristic 
of optically thin dust disks  (see Fig.~\ref{figexcess1} and \ref{figexcess2}). 
We use HD~37484 and its disk as our demonstrative case because the system
has the largest warm excess emission and the star has a low X-ray luminosity, 
thus representing the case with the largest 
dust surface density and the lowest X-ray heating of the gas. 
We assume that the disk has a dust surface density proportional to  $r^{-1}$
(e.g. the Vega best fit model to its  debris disk by \citealt{2005ApJ...628..487S}).  
 We estimate in three steps  the maximum dust surface density at the disk inner radius 
 that is consistent with the infrared observations:
i) we calculate the dust temperature using the analytical solution of the 
radiative transfer equation 
for optically thin disks (e.g. \citealt{2004A&A...417..793P}, eq. 5);
ii) we determine the excess emission by subtracting the best-fit Kurucz model atmosphere 
from the observed data;
iii) we write the dust re-emission in terms of the dust surface density at the 
inner radius (e.g. \citealt{1999ApJ...527..918W} eq. 3)
 and require that the re-emission does not exceed the observed excess at any wavelength.
The maximum dust surface density for inner radii from 1 to 19\,AU is shown in 
Fig.~\ref{figsurf} 
for a disk with small (0.2\,\micron) and large (2\,\micron) grains. 
The dust surface density for small grains is lower than that for large 
grains at all disk radii. 
This is because smaller grains are warmer than larger ones at the same distance 
from the star and because, for the same dust surface density, smaller grains have more total 
surface area than larger grains.

On the other hand, a lower limit to the dust surface density
is obtained if the dust is to dominate the heating by X rays.
Heating from gas-dust collisions is proportional to the  
product of the densities of dust and gas while X-ray heating 
is only proportional to the density of gas.
By equating the  contribution from gas--dust collisional heating and X-ray heating (see GH05 for
 the analytic forms of these heating processes), we obtain a 
 minimum dust density 
for dust-gas collisions to contribute as much as X-ray heating. 
We compare the two heating mechanisms at the
midplane of the disk where the number density is highest at any
given radius, and hence where collisional heating is maximum. This
is also the region where X-ray heating is at its lowest, since the
attenuation is highest. 
To convert the number density to surface density we use the mass of  spherical Draine \& 
Lee silicate grains
and a disk scale height of 0.1  times the radial distance, which is a good 
approximation for the inner regions 
of flared disks (e.g. \citealt{1997ApJ...490..368C}). 
The resulting minimum dust surface densities for different inner disk radii and for 
two grain sizes are summarized in Fig.~\ref{figsurf}. 
These values are obtained for a gas temperature of 70\,K. Note that
the general trend will not change if a different gas temperature is assumed, 
but
for gas hotter than the dust at any radius gas-dust collisions will result in 
cooling rather than heating the gas.
However, the magnitude of the collisional cooling is similar to the magnitude of 
the collisional heating,
the energy transfer being proportional to ($T_{\rm gas} - T_{\rm dust}$).  

The comparison of the two surface densities proves that our dust disks are too 
tenuous for dust to appreciably affect gas heating or cooling. 
This simple approach shows that at least two orders of magnitude higher dust surface 
densities are necessary for dust-gas collisions to equal X-ray heating. 
Because the other gas sources have lower excess 
emission than HD~37484 and/or higher X-ray luminosities, this result extends to 
 the entire sample. 
We also point out that grain photoelectric heating is completely negligible for
optically thin dust disks where most of the dust grains have grown to sizes larger
than the very small ($<< 0.2$\,\micron ) interstellar grains which dominate grain 
photoelectric heating (see the case of HD~105 by H05 and Fig.~3 by GH04).
We therefore proceed to model the gas temperature and density structure of disks 
ignoring their  dust properties.

\begin{table*}[htbp]
\begin{center}
\caption{Summary of UV luminosities and templates.}\label{tabUVtemp}
\begin{tabular}{lc | lcc |  c}
\tableline\tableline
\noalign{\smallskip}
\multicolumn{2}{c|}{FEPS sources} & \multicolumn{3}{c|}{UV templates}   &  \\
Source & Log($\frac{{L_{\rm UV}}}{{L_{\star}}}$) & Source & SpT & Criteria & Ref.\\
\noalign{\smallskip}
\hline
ScoPMS~214&  -2.2  & ScoPMS~52$^a$ & K0IV &   age, colors & 1 \\
MML~17    &  -2.4  & HD~146516$^a$ & G0IV &   age, colors & 1 \\
MML~28    &  -2.6  & ScoPMS~52$^a$ & K0IV &   age, colors & 1\\
HD~37484  &  -2.4  & HD~28568 & F2V   &   X-ray       & 2\\
HD~202917 &  -2.6  & HD~43162 & G5V   &  X-ray        & 2	\\
HD~134319 &  -3.2  & HD~134319     &	 &	      & 3  \\
HD~12039  &  -2.8  & HD~43162 & G5V   &  X-ray        & 2 \\
V343~Nor  &  -3.3  & V343~Nor	 &    & 	      & 3 \\
HD~377    &  -3.7  & EK~Dra & G1.5V   &  age, X-ray   & 4 \\
AO~Men    &  -2.5  & RE~J0137+18A& K3V&   age, colors & 3   \\  
HD~209253 &  -2.8  & HD~33262 & F7V & X-ray	      & 2\\
HD~35850  &  -2.8  & HD~35850	 &    & 	      & 3\\
HD~25457  &  -3.1  & HD~28033 & F8V   &  X-ray        & 2   \\
HD~17925  &  -3.5  & HD~17925	   &	 &	      & 3\\
HD~216803 &  -2.2  & HD~216803     &	&	      & 3  \\
\tableline\tableline
\tablecomments{
The UV luminosity we report here is calculated between 6.0 and 13.6\,eV for comparison
to the interstellar radiation field \citep{1969BAN....20..177H}.
}
\tablenotetext{a}{These two template stars belong to the Upper Scorpius association (d=145\,pc) and
have large visual extinctions as estimated from fitting their photometric data with Kurucz 
stellar atmospheres ($A_{\rm V}$=1.5 for
ScoPMS~52 and $A_{\rm V}$=0.8 for HD~146516). Their IUE spectra have been deredden using 
the reddening law by \citet{1990ARA&A..28...37M} before scaling them to the Kurucz model 
atmospheres of our targets. All other template stars have distances less than 65\,pc and thus 
negligible extinction: 
their IUE spectra scaled to the source distances and radii were found to match well 
the source Kurucz model atmospheres at wavelengths longward of 3200\,\AA.}
\tablerefs{1. \citet{2000ApJS..129..399V}; 2. \citet{2005ApJS..159..118W}
3. IUE archive at http://archive.stsci.edu/iue/ ; 4. \citet{2005ApJ...622..680R} }
\end{tabular}
\end{center}
\end{table*}

\subsection{Gas Models for Optically Thin Disks}\label{gasopthin}
The simple approximations adopted in Sections~\ref{upIRS} and \ref{upCO}  
allowed us to obtain initial estimates of gas mass upper limits and compare 
them to values published in the literature. The upper limits were shown to depend
strongly on the assumed gas temperature. We can improve on these estimates by applying comprehensive gas models to calculate the gas disk
temperature and thus infer {\it total} gas mass upper limits for the targeted disks.  We use the
models of GH04 that include the chemistry and thermal balance in a self--consistent manner and
calculate the vertical density structure and the temperature in the radial and vertical directions.
These models include heating and cooling from gas--dust collisions, stellar and interstellar UV
radiation, stellar X--rays, collisional deexcitation of vibrationally excited H$_2$ molecules,
grain photoelectric heating, exothermic chemical processes, and cosmic rays. 
Condensation of molecules onto grains is not implemented for two main reasons. First, 
infrared lines originate from regions where dust is too hot ($\ge$100\,K) for 
condensation to occur appreciably. 
Second, the code is steady state but would require time dependent solutions to 
properly treat the condensation of molecules. Gas--phase species are assumed to have interstellar
abundances (\citealt{1996ARA&A..34..279S} and Appendix~A in GH04). We demonstrated in Section~\ref{s_thindust}  that gas-dust collisions and grain photoelectric heating are insignificant heating 
and cooling mechanisms in tenuous dust disks and therefore we removed the dust component
in the models that follow.

{FIDUCIAL DISK MODEL.} We first consider a fiducial  face--on gas disk extending from 1 to
100\,AU with a surface density dependence of  $\Sigma\propto r^{-1}$, as
indicated from observations of young disks (e.g.
\citealt{1996A&A...309..493D}).  The main input parameters to the gas models are
the stellar properties reported in Table~\ref{stars} and stellar FUV fluxes
(UV fluxes between 6\,eV\,$<\, h \nu\,<$\, 13.6\,eV). 
Our first modeling of the disk around HD~105 showed that stellar X ray and UV
photons dominate the gas heating in tenuous dust disks (the UV heats not
by the grain photoelectric heating mechanism but by pumping H$_2$ and by photodissociating 
and photoionizing molecules and atoms, H05). Because stellar
X-ray and UV fluxes depend on the  stellar activity, we tried to collect
measurements for each source. X-ray data are available for all targets (see
Table~\ref{stars}) while only 5 of our sources have been observed in the UV by
the International Ultraviolet Explorer (IUE). For these sources we merged the IUE
data with the Kurucz model atmospheres\footnote{The Kurucz model atmospheres for
the stellar parameters in Table~\ref{stars}
match in all cases the long wavelength IUE data (from $\sim$3000\,\AA).
We used the IUE data for wavelengths $<$3000\,\AA{} and the Kurucz model atmospheres for
longer wavelengths.} 
and estimated the flux in 8 energy bins
(from 0.7 to 13.6\,eV) that are relevant for the photodissociation and
photoionization of dominant atomic and molecular species like S, Fe, Mg, Si, CO
and H$_2$. Where we lack UV data for our targets, we searched for stars with
similar spectral type, age, and colors in the IUE archive and used them as
templates for the FUV flux. For sources where ages are not well-determined we
used the spectral type and X--ray luminosity as the main criteria for assigning
templates. We estimated the H~I Ly-alpha flux from the  X-ray flux, based on
the empirical relationship by \citet{2005ApJS..159..118W}.  
Since Ly-alpha emission contributes most of the FUV emission from late-type stars, 
we use it as a proxy for the total FUV emission and scale the templates accordingly, 
yielding order-of-magnitude estimates for the stellar UV field. A
summary of UV templates and luminosities is given in Table~\ref{tabUVtemp}.  In
Sect.~\ref{mod-dependence} we will discuss the dependence of our results on
the input parameters that are most uncertain: the disk inner radius, the surface density
slope, and the stellar UV luminosity.

{METHOD.}  We calculate the line fluxes from gas models with the total disk mass
as our main variable parameter. For each transition reported in
Tables~\ref{line-fluxes} and \ref{tab:upCO}, we find a disk mass where the
calculated line flux from the models matches the observed flux limit. We
determine the radius, $R_{\rm em}$, from within which 90\% of the emission
originates for each transition and also define a disk mass associated with this
region $M_{\rm em}$, which includes mass from the inner radius to $R_{\rm em}$. We also calculate
the radius $R_{\rm thick}$ within which the CO emission is optically thick. This
region corresponds to a  vertical column density of CO molecules of $10^{15}-10^{16}$\,cm$^{-2}$ (this value is similar for the two CO
lines and corresponds to line optical depths larger than 1). Our modeling results
are summarized in Tables~\ref{gas_mod} and \ref{gas_mod_co}.

{RESULTS FROM THE MID--INFRARED LINES. }  We find that the \si{} transition at
25.23\,\micron{} sets the  most stringent upper limits on the gas mass in
comparison to the limits set by the other mid--infrared lines. 
 We explored the sensitivity of our results to the S abundance in the gas phase
using the  disk of HD~35850 as a test case. Observations from
comet Halley dust suggest that about half of the S could be in FeS grains 
(e.g. \citealt{1994ApJ...421..615P}). Some S could be removed from the gas phase component of circumstellar disks in this way, as recently proposed by \citet{2002Natur.417..148K}. However, current observations cannot quantify the amount of S into grains.  To investigate a pessimistic case we reduced the S abundance in gas phase
by  90\%.  This reduction results in a 5 times larger gas mass upper limit. This is because for such a low mass  disk the \si{} line is nearly optically thin (optical depths of 2--3), so that the \si{} 
flux drops nearly proportional to the gas mass. Note that even if  \si{} upper limits  would be increased by a factor of 5, they would still be more stringent that those inferred from H$_2$ lines for most of the sources (see below). 

Our gas models predict also relatively strong \SiII{} lines at 34.8\,\micron , however  the IRS spectra
become noisier at longer wavelengths and our line flux upper limits are not
as low. These results are in agreement with what we found for the disk around
HD~105 (H05). The \feii{} line at 26\,\micron{} is the second most sensitive
transition and can set  useful gas mass upper limits  for two-thirds of the sample (in
other words, in one-third of the sample, our models did not produce  detectable
\feii{} no matter how much gas mass was contained within the disk).  The H$_2$
S(1) line at 17\,\micron{} provides the most stringent upper limits when
compared to the other two  H$_2$ lines; nevertheless gas masses can be
constrained only for one-third of the sample with this diagnostic.
This is because the transition
probabilities of  the H$_2$ lines are orders of magnitudes smaller than those of the 
\si{} and \feii{}  transitions while  their excitation temperatures are quite similar.
The H$_2$ surface density upper limits are typically a factor of 10 higher than
the surface densities  set by \si , while the \feii{} transitions provide values
similar to those from \si{} for many sources. 

For an inner disk radius of 1\,AU, the models indicate that the emitting region for the 
mid--infrared transitions is typically  a few AU  and extends up to $\sim$5\,AU in a few of our fiducial disks,  suggesting that we are tracing a region analogous to that between Earth and Jupiter in our Solar System.  
We note that the  gas mass upper limits inferred from modeling the \si{}
lines are less than 0.4\,$M_\oplus$ in the 1-5\,AU region   for all sources and
are as low as a tenth of an Earth mass for two--thirds of the sample.  

 Comparing gas masses from the models to the simple LTE approximation (Sect.~\ref{upIRS})
requires knowing the characteristic temperature of the emitting gas. For the 5 sources
where we could set limits from the   H$_2$~S(1) line, we computed mean temperatures
as follows: $$ <T> = \sum F_i \times T_i /  \sum F_i \,\, ,$$ where the sum is carried out over the spatial grid
and the flux is the H$_2$~S(1) flux. We obtain mean temperatures between 230\,K
(for HD~35850) and 160\,K (for HD~216803). This temperature range reflects the different 
source heating;  K-type stars like HD~17925 should have even lower $<T>$ 
close to 100\,K. 
Limits on the mass of the emitting  gas from modeling the \si{} line ($M_{\rm em}$ in Table~\ref{gas_mod})  are typically a factor of 50 and of 10 lower than the warm gas limits calculated in Sect.~\ref{upIRS} for gas temperatures
of 150\,K and 200\,K respectively. The difference results from the 
use of \si{} as  the tracer and the temperature structure in real disks.  The limits on the {\it total} gas mass in a disk depends on the power law of the
gas surface density and on the outer disk radius.  For our fiducial disk model
($\Sigma \propto r^{-1}$, $1\le r \le$\,100\,AU)
upper limits to the total gas mass range from 10\,$M_\oplus$ (for ScoPMS~214)
to   0.5\,$M_\oplus$ (for HD~17925), with higher limits typically for the 
more distant sources, which are also the younger 
(see last column of Table~\ref{gas_mod}).

\begin{figure*}
 \resizebox{\textwidth}{!}{\includegraphics[angle=0]{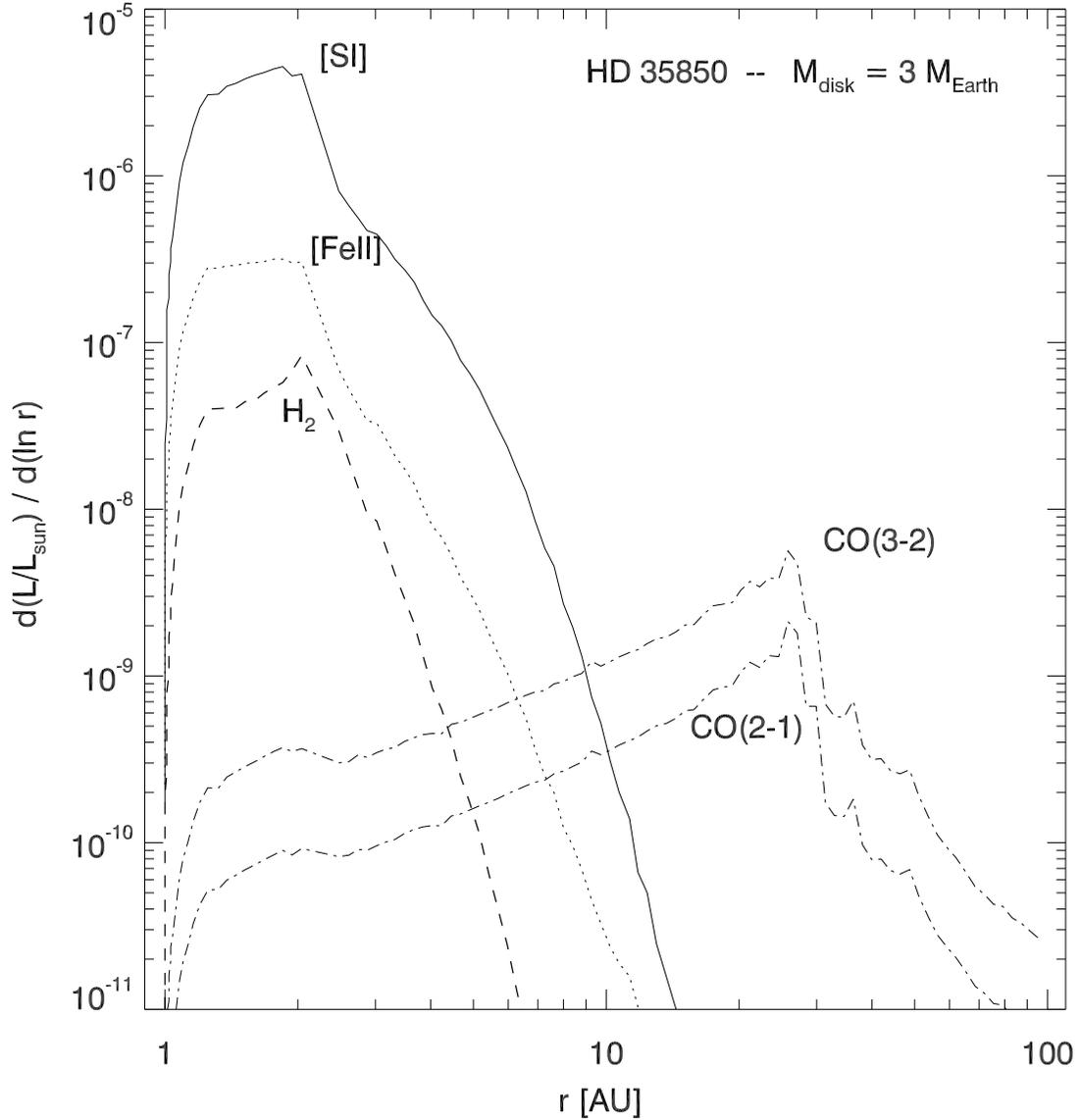}}
\caption{Emitting regions for mid-infrared and millimeter transitions.
Plotted is the increment in line luminosity within a logarithmic radial 
annulus versus the disk radius. The luminosities of the H$_2$ (dashed line), 
\feii{} (dotted line), and \si{} (solid line) transitions steeply increase 
at the disk inner radius, remain high out to few AU, and rapidly decline  at larger radii. 
The luminosities of millimeter CO transitions (dash-dotted lines) increase 
more gradually with radius.
The drop in luminosity at $\sim$30\,AU occurs because CO becomes optically thin 
 to its own radiation at  these  radii and is rapidly photodissociated. 
The wiggles are due to the coarse grid spacing used for the calculation in the outer regions.
}
\label{fig_em}
\end{figure*}

{RESULTS FROM THE MILLIMETER CO LINES. } 
While mid-infrared line luminosities trace the inner few AU of circumstellar
disks,  millimeter CO transitions are sensitive to colder gas located in the
outer disks. CO is mainly heated by stellar X-ray and UV emission, 
but at large disk radii 
($\sim$20\,AU for a star with $L_{\rm UV}\sim 10^{-3}\,L_\sun$) 
the interstellar UV  field can dominate the local stellar UV.
 Fig.~\ref{fig_em} shows the line 
luminosity per logarithmic radial annulus produced in a disk as a function 
of $r$. Mid-infrared line luminosities increase steeply at the disk inner 
radius, peak in the inner few AU, and rapidly fall as the temperature decreases.
In contrast, CO line luminosities gradually increase with radius, peak at 
larger disk radii, and decline as CO begins to photodissociate.
CO emission is so widespread that radii as large as 20\,AU and up to
60\,AU (see $R_{\rm em}$ of Table~\ref{gas_mod_co}) are required to encompass
90\% of the CO luminosity. This emission is mostly optically thick, suggesting
that as a first approximation CO is mainly  tracing the radius inside of which the
CO is optically thick rather than the total disk mass.
The disk radii estimated with the simple assumption of optically thick CO  emission 
and 20\,K temperature are typically a factor of 2 higher (and can be up to a factor of 4 higher, e.g.  HD~12039) than the $R_{\rm thick}$ from the models. This suggests that gas temperatures as high as
$\sim$50\,K are more representative of the region where CO is optically thick
(see also the discussion in Sect.~\ref{upCO}).
 In the case of HD~17925  the low observed flux limits only
allow a very tenuous gas disk. The low column density of
the disk results in photodissociation of CO by the interstellar
UV field (the average H$_2$/CO number ratio is 10$^5$, 
an order of magnitude higher than the interstellar value) 
and CO emission from this disk is optically thin  at all radii. 

\begin{table*}
\begin{center}
\caption{Results derived from the Spitzer observations and the gas models.}
\begin{tabular}{l |ccc| ccc| ccc | c}
\tableline\tableline  
\noalign{\smallskip}
Source&\multicolumn{3}{|c}{$\Sigma_0$\,(g/cm$^2$)} & 
\multicolumn{3}{|c}{$R_{\rm em}$\,(AU)} & \multicolumn{3}{|c|}{$M_{\rm em}$\,($M_\oplus$)} & $M_{\rm disk}$  \\

&  \mh & \feii & \si & \mh & \feii & \si  & \mh & \feii & \si  & ($M_\oplus$)  \\
\noalign{\smallskip}
\tableline	 
ScoPMS~214 &  ...   &  ...  & 0.41 &   ... & ...   &3.4   & ...  & ...   & 0.23  &9.6  \\ 
MML~17	   &  ...   &  ...  & 0.29 &   ... & ...   &2.9   & ...  & ...   & 0.13  &6.8  \\
MML~28	   &  ...   &  ...  & 0.13 &   ... & ...   &2.4   & ...  & ...   & 0.04  &3.0  \\
HD~37484   &  ...   & 0.68  & 0.13 &   ... & 2.3   &1.6   & ...  & 0.21  & 0.02  &3.0   \\
HD~202917  &  ...   & 0.48  & 0.26 &   ... & 2.0   &2.9   & ...  & 0.11  & 0.12  &6.1   \\
HD~134319  &  ...   &  ...  & 0.13 &   ... & ...   &1.7   & ...  & ...   & 0.02  &3.0   \\
HD~12039   &  0.7   & 0.12  & 0.25 &   2.8 & 3.4   &2.6   & 0.3  & 0.07  & 0.09  &5.8  \\
V343~Nor   &  ...   & 0.05  & 0.09 &   ... & 4.6   &5.1   & ...  & 0.04  & 0.09  &2.1  \\
HD~377	   &  1.3   &  ...  & 0.12 &   1.4 & ...   &1.6   & 0.1  & ...   & 0.02  &2.8   \\
AO~Men	   &  0.9   & 0.13  & 0.09 &   4.2 & 2.7   &3.3   & 0.7  & 0.05  & 0.05  &2.1  \\
HD~209253  &  ...   & 0.29  & 0.06 &   ... & 1.8   &2.5   & ...  & 0.06  & 0.02  &1.4    \\
HD~35850   &  1.2   & 0.04  & 0.11 &   2.5 & 4.2   &5.5   & 0.4  & 0.03  & 0.12  &2.6  \\
HD~25457   &  ...   & 0.41  & 0.07 &   ... & 1.7   &2.9   & ...  & 0.07  & 0.03  &1.6   \\
HD~17925   &  ...   & 0.36  & 0.02 &   ... & 1.4   &1.8   & ...  & 0.03  & 0.004 &0.5    \\
HD~216803  &  0.3   & 0.07  & 0.08 &  2.5  & 1.5   &1.9   & 0.1  & 0.008 & 0.02  &1.9   \\
\tableline
\end{tabular}\label{gas_mod}
\tablecomments{
Upper limits to the gas obtained from the line flux upper limits of \mh{} at 17\,\micron , 
\feii{} at 26\,\micron , and \si{} at 25.23\,\micron{} and our fiducial disk model.
$\Sigma_0$ is the gas surface density at the disk inner radius  (1\,AU).
$R_{\rm em}$ is the radius within which 90\% of the emission originates.
$M_{\rm em}$ is the gas mass between 1\,AU and $R_{\rm em}$.
Total disk masses from the \si{} upper limits (last column) and our fiducial disk range from
0.5 (for HD~17925) to 8 (for ScoPMS~214) $M_\oplus$. However, note that they are not
tightly constrained by the Spitzer observations alone, which are only sensitive to gas 
between the inner disk radius and $R_{\rm em}$. }
\end{center}
\end{table*}

\begin{table*}
\begin{center}
\caption{Results derived from the SMT observations and the gas models.}
\begin{tabular}{l| cc | cc | cc | c}
\tableline\tableline
\noalign{\smallskip}
Source  &\multicolumn{2}{|c}{$R_{\rm em}$ (AU)} & \multicolumn{2}{|c|}{$M_{\rm em}$ ($M_\oplus$)} 
        &\multicolumn{2}{|c|}{ $R_{\rm thick}$ (AU)} & $M_{\rm disk}$ ($M_\oplus$)\\
       & CO(2-1) & CO(3-2) & CO(2-1) & CO(3-2) & CO(2-1) & CO(3-2) & CO(3-2) \\
\tableline	 
\noalign{\smallskip}
ScoPMS~214  & ...  & ...  & ... & ...   & $>100^{(100\%)}$  & $>100^{(100\%)}$  & 17\\
HD~37484    & 49.8 &  --  & 3.5 &  --  & 50.1$^{(90\%)}$ & --                   & 7.0\\
HD~134319   & 50.4 & 61.3 & 1.6 &  1.6 & 43.5$^{(87\%)}$ & 43.5$^{(83\%)}$      & 2.7\\
HD~12039    & 30.4 & 31.9 & 1.0 &  0.9 & 20.5$^{(62\%)}$ & 19.6$^{(56\%)}$      & 2.8\\
HD~377	    & 36.4 & 53.8 & 1.2 &  2.3 & 33.0$^{(86\%)}$ & 44.5$^{(84\%)}$      & 4.4 \\
HD~209253   &  --  & 31.1 &  -- &  0.7 & --              & 28.2$^{(85\%)}$      & 2.6\\
HD~35850    & 23.0 & 21.9 & 0.7 &  0.6 & 20.9$^{(87\%)}$ & 13.4$^{(50\%)}$      & 2.7\\
HD~25457    & 28.4 & 26.4 & 0.5 &  0.4 & 28.2$^{(91\%)}$ & 28.2$^{(95\%)}$      & 1.5\\
HD~17925    & 21.1 & 20.1 & 0.07&  0.05& 0$^{(0\%)}$ & 0$^{(0\%)}$              & 0.3\\
\tableline
\end{tabular}\label{gas_mod_co}
\tablecomments{
Upper limits to the gas obtained from the SMT observations.
$R_{\rm em}$ is the radius within which 90\% of the emission originates.
$M_{\rm em}$ is the mass between 1\,AU and $R_{\rm em}$.
The CO(2-1) and CO(3-2) transitions trace colder gas in comparison to mid-infrared lines 
and thus are sensitive to the outer disk regions.
$R_{\rm thick}$ is the radius at which the CO line optical depth is equal to 1,
corresponding to a column density of CO molecules (perpendicular to the disk,
along the vertical direction) approximately equal to $10^{15}\,cm^{-2}$.
 The number in parenthesis gives the percentage of optically
thick emission in comparison to the total CO emission. The last column gives the total disk (hydrogenic) 
masses from 1 to  100\,AU for our fiducial disk surface density. 
 These values are computed from the CO(3-2) transitions for all sources except ScoPMS~214
and HD~37484 where we used CO(2-1) data. In the case of ScoPMS~214 the CO(3-2) line flux
upper limits are not stringent enough to set useful gas mass upper limits. 
For this gas mass the CO emission is optically thick all the way out to 100\,AU. In the case of HD~37484 we have 
only CO(2-1) data}.
\end{center}
\end{table*}

\subsection{Dependence of the Results on Uncertain Input Parameters}\label{mod-dependence}
In this section we test the dependence of our results on the 
slope of the gas surface density, on the stellar UV field, and on the 
disk inner radius, that are the main input parameters to our gas models lacking 
independent observational constraints.

\begin{figure*}
\resizebox{\textwidth}{!}{\includegraphics[angle=0]{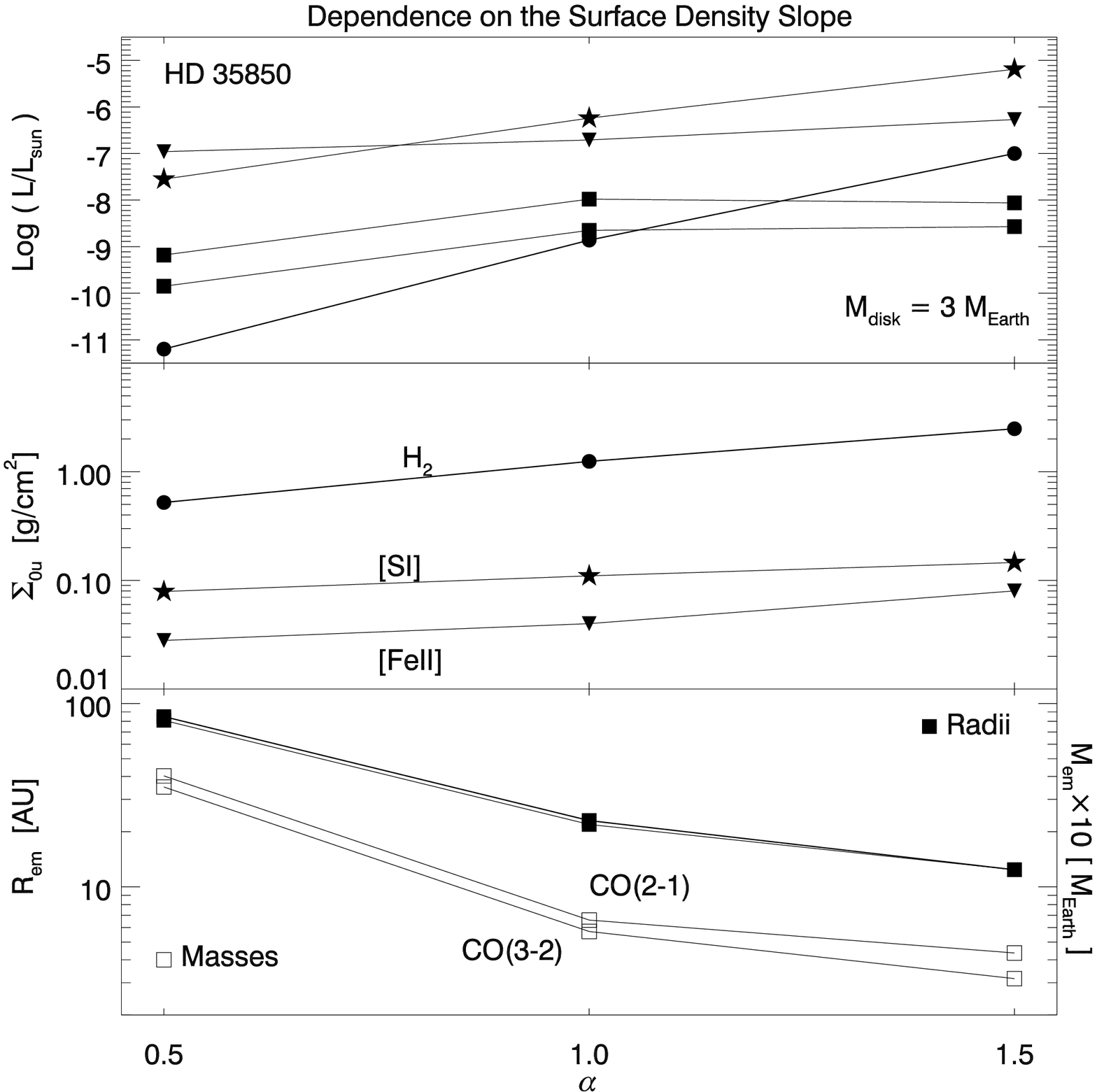}}
\caption{Dependence of our results on the surface density power law 
($\Sigma \propto r^{-\alpha}$).  
Changes in line luminosities for a disk of 3$M_{\oplus}$ around HD~35850 
are plotted in  the upper panel.
Stars, upside down triangles, circles, and squares denote lines luminosities for the \si{}, \feii{}, H$_2$~S(1), and  CO transitions respectively.
The middle panel shows the variation in the surface density upper limits 
at 1\,AU ($\Sigma_{0u}$).  
Steeper density slopes ($\alpha=1.5$) increase the mid--infrared line 
luminosities and lower the total gas mass upper limits. 
However, the surface density at the inner radius also increases with alpha for a given disk mass, that's why the limits on the surface density are higher for $\alpha = 1.5$  (see also text).
The extension (filled squares) and mass (open squares) of the region 
emitting in the millimeter is plotted in the lower panel.}
\label{fig_sden}
\end{figure*}

First, we consider different slopes of the gas surface density
power-law, $\Sigma(r)\propto r^{-\alpha}$, and use the disk around
HD~35850 as the demonstrative case. In addition to our fiducial disk
model with $\alpha=1$, we summarize in Fig.~\ref{fig_sden}  results for a
flatter ($\alpha=0.5$) and a steeper ($\alpha=1.5$) surface density
distribution, thus comprising  the observed range of surface density slopes in
circumstellar disks. For a disk with fixed mass, distributing more mass in 
the inner regions ($\alpha=1.5$) results in increased mid-infrared line 
luminosities (upper panel of Fig.~\ref{fig_sden}). 
But a steeper density slope reduces the 
radius $R_{\rm em}$ of the emitting region and results in
higher surface density upper limits  at 1\,AU (middle panel of
Fig.~\ref{fig_sden}).   For instance, the $R_{\rm em}$ for the \si{} line
decreases from 10.4\,AU for $\alpha=0.5$ to 3.7\,AU for $\alpha=1.5$, with the
intermediate value of 5.5\,AU for our fiducial disk (see Table~\ref{gas_mod}).
The H$_2$~S(1) transition is more sensitive to the redistribution of mass than the
\feii{} and \si{} lines because it is the least optically thick transition.
Differences between  the limiting surface densities are less than a factor of 2 for
$\alpha$ differing by 0.5 (middle panel of Fig.~\ref{fig_sden}). CO lines are
not affected appreciably for $\alpha=1$ or larger but distributing more mass in
the outer regions ($\alpha=0.5$) results in a more extended CO emission and
therefore an increase of the mass associated with the emission (by a factor of
6). However, the outer disk mass between two fixed radii (in the zone
10--50\,AU) differ by less than a factor of 2. Thus we conclude that
uncertainties in the surface  density slope introduce only a factor of a few
uncertainty  in our results.

\begin{figure*}
 \resizebox{\textwidth}{!}{\includegraphics[angle=0]{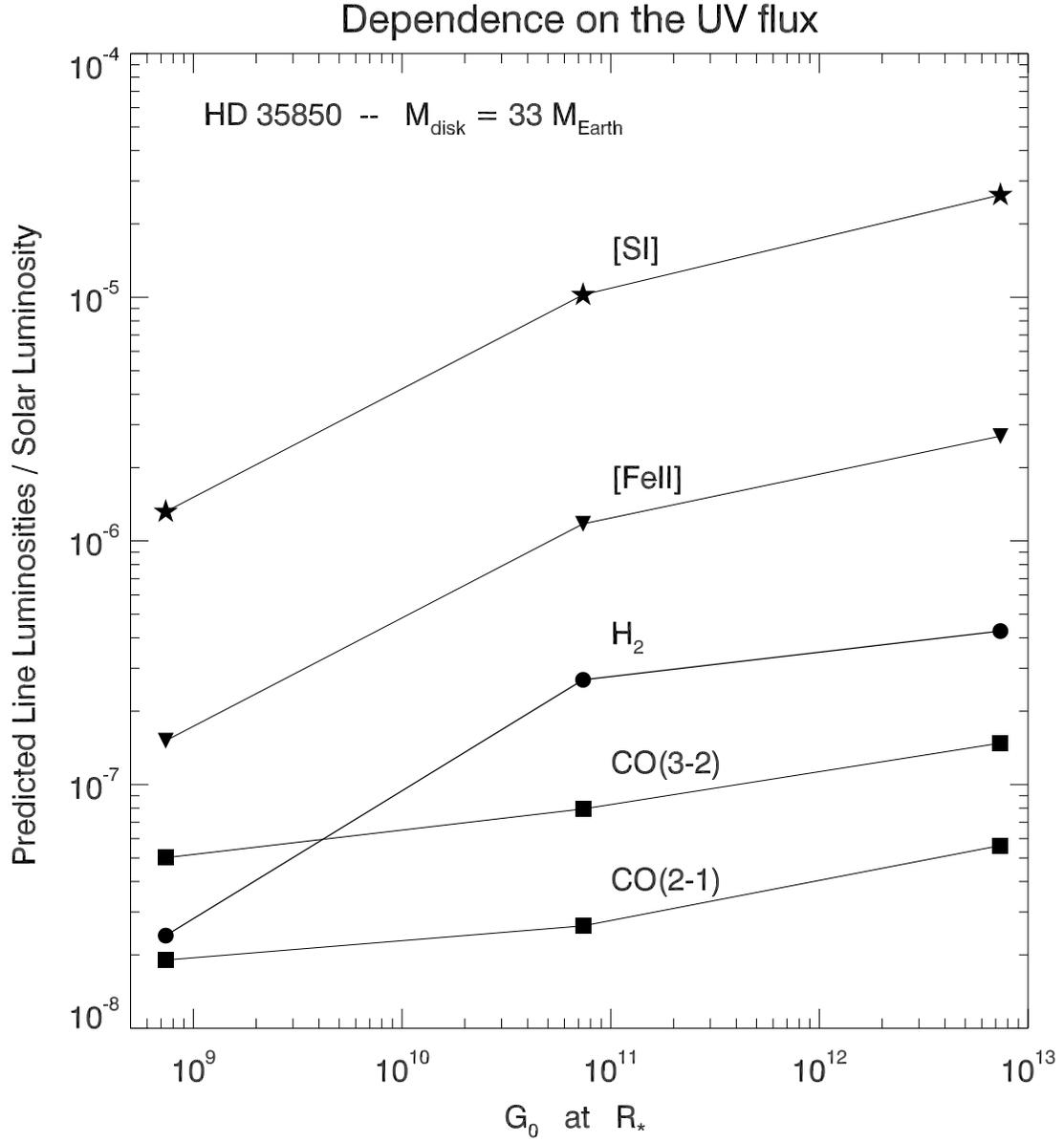}}
\caption{Dependence of line luminosities on the UV flux (see Fig.~\ref{fig_sden} for the symbols).
These luminosities have been computed for a disk of 33$M_{\oplus}$ 
(from 1 to 100\,AU) around the star HD~35850. $G_0$ is the UV flux between 6 and 13.6\,eV at the stellar
surface normalized to the interstellar "Habing flux" 
(1.6$\times10^{-3}$\, erg\,cm$^{-2}$\,s$^{-1}$).}
\label{fig_uv}
\end{figure*}

We then consider the dependence of our results on  different stellar UV luminosities.  In addition to the UV flux for HD~35850  ($G_0=7.37\times10^{10}$ at the stellar radius\footnote{$G_0$ is the
flux between 912-2000\AA{} in units of 1.6$\times10^{-3}$\,erg\,cm$^{-2}$\,s$^{-1}$})  assumed previously,  we have modeled two cases with 100 times higher and 100 times lower UV flux. 
The effect of higher stellar UV flux is to increase the heating (as well as  alter the chemistry) 
and thus the predicted line luminosities of mid-infrared as well as of millimeter lines (see
Fig.~\ref{fig_uv}). Note that an increase of 4 orders of magnitude in the UV flux
increases the mid--infrared and millimeter line luminosities by factors smaller
than 20 and 3 respectively. Because our method provides order--of--magnitude
estimates of the stellar UV field, model line luminosities have only a factor of a 
few uncertainty.  Based on the previous tests on the surface density slope,  such luminosity uncertainties  
should have negligible effects on the limiting surface densities and on the extension of the emitting 
regions (compare Figs.~\ref{fig_sden} and \ref{fig_uv}).

Finally, we test the dependence of our results on the disk inner radius. Our  earlier
modeling of HD~105 (H05) showed that, for a range of $R_{\rm in}$ between 1\,AU and
40\,AU,   the upper limit on the surface density at the inner radius  varied only by
a factor of a few. This result is robust and can be extended to our sample:  {\it Spitzer} data
can set useful upper limits on the surface density for disks with $R_{\rm in}$ out to 40\,AU\footnote{Note that $R_{\rm em}$ will change depending on  $R_{\rm in}$. The models from H05 suggest that  the extension of the region where {\it Spitzer} lines originate is a few AU  
(see Table 2 from H05, there we used the notation $r_{\rm w}$ instead of $R_{\rm em}$)}. 
For inner radii smaller than 1\,AU,
H05 found that the surface  densities required
to produce detectable lines increased (see their Fig.~6).  This is because for
smaller inner radii the mass or area   (when lines are optically thick) of the emitting
gas decreases.  Essentially all the heating is deposited close to the star and gas located at
larger radii becomes too cool to produce any line emission at mid--infrared wavelengths.
 The result was that if the disk inner radius of HD~105 became
smaller than  $\sim 0.5$\,AU, line upper limits from {\it Spitzer}  could not set
useful constraints on  the gas mass. 

\begin{figure*}
 \resizebox{\textwidth}{!}{\includegraphics[angle=0]{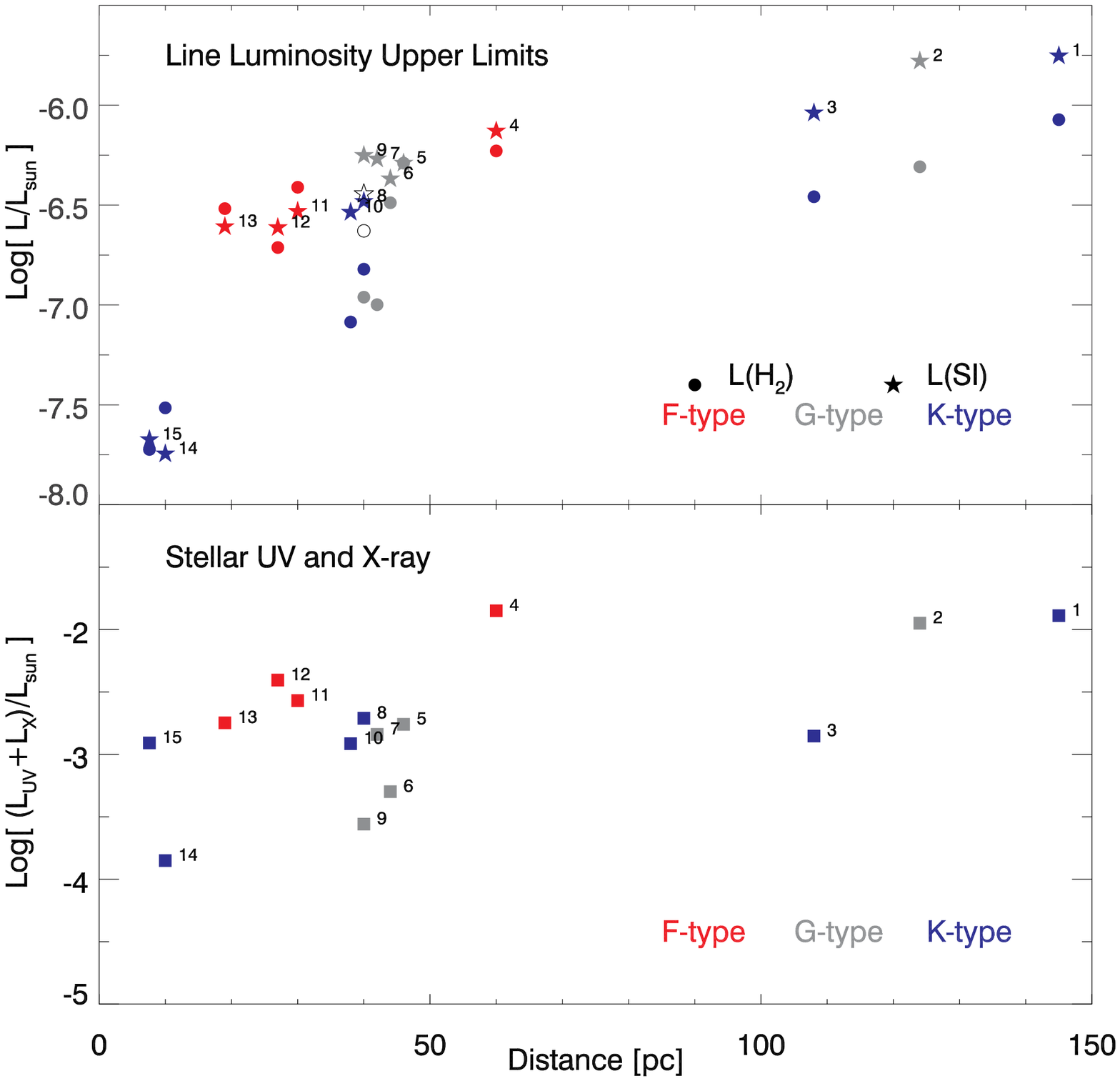}}
\caption{Upper panel. {\it Spitzer} line luminosity upper limits versus source 
distance. Stars and circles are for the \si{} and H$_2$~S(1) transitions respectively. 
For comparison we report with open symbols the upper limits to HD~105 (H05). 
Symbols in red, gray, and blue indicate values from F-, G-, and K-type stars.
Line luminosity limits are higher for the more distant sources.
Lower Panel. Stellar UV and X-ray luminosities versus source distance.
Early-type stars and young/more distant sources have higher luminosities.
}
\label{figLineup}
\end{figure*}

\begin{figure*}
 \resizebox{\textwidth}{!}{\includegraphics[angle=0]{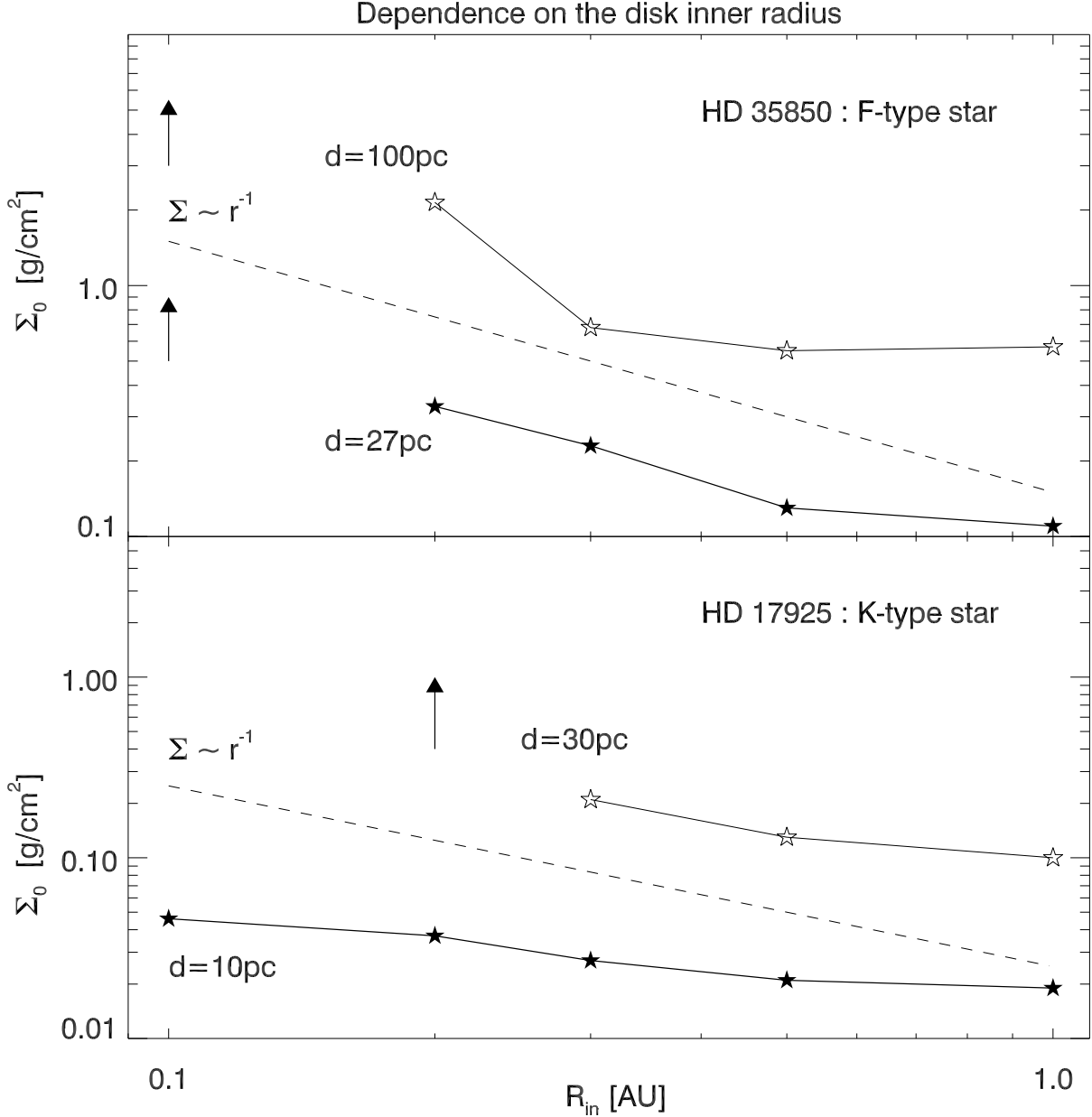}}
\caption{Dependence of the gas surface density upper limits ($\Sigma_0$) 
on the disk inner radius. 
$\Sigma_0$ is calculated from the upper limits to the \si{} line at the 
source distance (filled symbols) and at 30\,pc and
100\,pc for HD~17925 and  HD~35850 respectively (open symbols). 
More distant sources have higher line luminosity upper limits
that result in larger upper limits $\Sigma_0$ and in larger $R_{\rm in}$ 
where the gas disk mass becomes unconstrained by {\it Spitzer} observations 
(this radius is marked by vertical arrows).
For sources at  a similar distance, upper limits to $\Sigma_0$ are set by the 
stellar X-ray and UV fields and more constraining limits are obtained for 
disks around early-type stars (compare the results for HD~35850 at 27\,pc and HD~17925 and 30\,pc).  
The dashed line gives the dependence of our fiducial surface density with radius. 
The rise in $\Sigma_0$ for $R_{\rm in}<0.5$\,AU
is steeper than the fiducial surface density slope for the more distant 
sources, implying somewhat higher surface density upper limits at 1\,AU for these disk inner radii.
}
\label{figRin}
\end{figure*}

In general, the sensitivity to the disk inner radius
depends on the heating  (stellar X-ray and UV) and on the line upper limits from
{\it Spitzer}.  While the younger sources in our sample typically have higher X-ray and UV
luminosities, and  therefore produce more line flux, they are also more distant and
so   the upper limits on their luminosities are less stringent (see Fig.~\ref{figLineup}). To
illustrate the effect of heating from stellar X-rays and UV we use the two disks  
around the F-type star HD~35850 and the K-type star HD~17925. We have chosen these
sources  because they represent extremes in stellar UV and X-ray luminosity
(HD~35850 has high UV and X-ray luminosities, whereas HD~17925 has low UV and X-ray luminosities) and have IUE
spectra and  CO millimeter data (sources 12 and 14 in Fig.~\ref{figLineup}).  We
consider 5 disk models with inner disk radii equal to  1, 0.5, 0.3, 0.2, and
0.1\,AU and compute for each case the surface density upper limits at  the inner
radius ($\Sigma_0$) from the \si{} transition  (filled symbols in
Fig.~\ref{figRin})\footnote{We have also computed the $R_{\rm em}$ from CO lines
and noted only marginal changes with $R_{\rm in}$}.  To show the effect of source
distance, we also compute upper limits for HD~17925 at 30\,pc 
(Log[L(SI)/L$_\sun$]= -6.8) and  HD~35850 at 100\,pc (Log[L(SI)/L$_\sun$]=-5.5), i.e. at about 3 times their real distance (open symbols in
Fig.~\ref{figRin}).  To calculate these line luminosities we have assumed that the flux limits and the S/N 
do not change with the source distance and simply scaled the luminosities with the distance square  
(in reality our limits from the \si{} line  are better for far--away sources due to longer exposure times 
but only by a factor of 4 at most).  These tests show that:  a) more distant sources will typically have larger
$\Sigma_0$ limits as well as larger disk inner radii at which the gas mass becomes
unconstrained by {\it Spitzer} observations; b) high stellar heating lowers the
limits on the gas surface density for sources at comparable distance  (i.e.,
comparing HD~35850 at 27\,pc and HD~17925 at 30\,pc we see that 
we can set useful limits until $R_{\rm in}$=0.2\,AU for the more luminous
HD~35850). Given the {\it Spitzer}
upper limits and the stellar properties of our sample (Fig.~\ref{figLineup}), we
conclude that we can set useful upper limits to the gas mass in the majority of our
systems for disks with inner holes ranging from $\sim$0.3 to 40\,AU. In the case of
the more distant sources  ScoPMS~214, MML~17, and MML~28, which are K- and G-type
stars, the inner radius at which the surface density cannot be constrained may be
larger than 0.3\,AU but still less than 1\,AU. 

\section{Discussion}\label{sect:discuss}
We have shown that infrared and millimeter observations set low limits
to the amount of gas in the inner ($\sim$1-5\,AU)  and outer
($\ge$15\,AU) regions, respectively, of our fiducial disks.    In
Sect.~\ref{mod-dependence} we have proven that order--of--magnitude
uncertainties in the stellar UV field and surface density slopes   varying
from -1.5 to -0.5 introduce only a factor of a few uncertainty in our
results. Reducing the disk inner radius does not affect appreciably the
result from the CO millimeter transitions because they originate farther
out in the disk. On the other hand, mid-infrared lines are sensitive to
small inner disk radii ($<$1\,AU). Depending on the stellar properties,
there is an inner radius at which {\it Spitzer} observations cannot set
useful upper limits to the gas mass because the area or mass of the
emitting gas is so small that lines cannot produce enough  luminosity
regardless of the surface density or the total disk mass. This critical
inner radius is $\lesssim$0.3\,AU for the majority of our targets, and
between 0.5 and 1\,AU for the three more distant systems (ScoPMS~214, MML~17, and MML~28). 
Thus the {\it Spitzer} observations alone can exclude the existence of gas disks with inner holes 
$\gtrsim 0.3$\,AU up to $\sim 40$\,AU around the nearby ($d\le 60$\,pc) 12 targets  
and gas disks with inner holes $\ge 1$\,AU up to $\sim 40$\,AU around the 3 targets 
at distances larger than 100\,pc.

We want now to convert surface density upper limits from the \si{} line, the 
most sensitive mid--infrared transition,  to gas mass upper limits 
and answer the question of whether the
targeted disks have enough gas at their present age to form gas giant planets.
For $R_{\rm in}\gtrsim$1\,AU, we have shown in the previous Section that the \si{} 
line traces the inner few AU (with 3\,AU being the mean value from Table~\ref{gas_mod}) 
and provides very similar surface density upper limits   at the disk inner radius. 
Thus, we can use the $\Sigma_0$ at 1\,AU in Table~\ref{gas_mod} as representative of 
the limiting gas surface density at the inner radius for disks with $R_{\rm in}$ between 1 and 40\,AU.  
In this way, we find  that {\it Spitzer} data can exclude gas masses larger than $\sim 0.04$\,M$_{\rm J}$ 
within 3\,AU of the inner radius for disks with inner radii between 1 and 40 \,AU.\footnote{These masses are not very sensitive to the assumed gas surface density slopes because of the small (few AU) region probed by mid--infrared gas lines.}  
This result is valid for all targets in our sample.

 In addition, for the 12 nearby systems we can compute limits on the gas mass for disks with inner radii 
smaller than 1\,AU. In the case of  the nearby source HD~17925 (only 10\,pc away) which has very stringent 
line flux upper limits, we can exclude gas disks with inner holes as small as 0.1\,AU and  a surface density at 
1\,AU even smaller than those reported in Table~\ref{gas_mod} (see Fig.~\ref{figRin}). 
This result can be extended to  HD~216803 which is similarly nearby and experiences even higher heating (Fig.~\ref{figLineup}). 
More generally,  limits on $\Sigma_0$ for $R_{\rm in}<$1\,AU increase faster than our assumed density slope and results 
in higher (up to a factor of 10 for $R_{\rm in}$ close to 0.3\,AU ) surface density upper limits than those reported in 
Table~\ref{gas_mod} (see Fig.~\ref{figRin} and the HD~105 models from H05).  However, even if assume that  $\Sigma_0$ is 
20 times larger  at 0.3\,AU (which gives 5\,g/cm$^2$), we find that  there is less than 0.0008\,M$_{\rm J}$ of gas between 
0.3 and 1\,AU; a constant surface density slope would have resulted in less than 0.002\,M$_{\rm J}$ of gas.

But could we also rule out a substantial amount ($\sim$M$_{\rm J}$) of gas within 1\,AU 
in the case of the three more distant systems?  A gaseous disk extending close to the star
may be accreting. We can use the upper limits on the stellar accretion rates for these sources to constrain 
the amount of gas in this disk region. The mass accretion rate of a
steady $\alpha$ disk around a solar-type star is\footnote{Eq.~\ref{eq_accretion}
was derived from eq. 5.78, 5.64 and 5.31 from \citet{1998apsf.book.....H},
assuming a gas surface density $\Sigma \propto r^{-1}$ and the isothermal sound 
speed for a hydrogen gas.  We normalize the temperature to 100\,K because that is close to the
dust temperature in the midplane of accreting T~Tauri disks.}:
\begin{equation}
 \dot{M} \simeq 7 \times 10^{-9} \, ( \frac{\alpha}{0.01} ) 
 \, ( \frac{T_{\rm 1AU}}{100K} ) \, ( \frac{\Sigma_{\rm 1AU}}{100\,g/cm^2} )   \, \, \,  [M_\sun/yr]
\label{eq_accretion}
\end{equation}
where $\alpha$ is the viscosity parameter (typically 0.01 from
magnetorotational instability models), and $T_{\rm 1AU}$ and $\Sigma_{\rm 1AU}$ are
the gas disk temperature and the surface density at 1\,AU. 
Accretion rates of $\sim 10^{-8}\,M_\sun/yr$ are  routinely measured in
disks surrounding young ($\sim$1\,Myr) classical T~Tauri stars by modeling the  UV
excess emission and the profile of the H$\alpha$ emission lines. 
\citet{2000ApJ...535L..47M} have extended these techniques to 10\,Myr old
stars in the TW~Hya association and detected accretion 
rates down to $\simeq 10^{-11}\,M_\sun/yr$ for lower mass stars than are studied
here. All our targets have been observed
in H$\alpha$ with high spectral resolution and found to have only narrow
H$\alpha$ lines (equivalent widths $<5\,\AA$) in absorption 
\citep{2005...white}.
The absence of H$\alpha$ in emission indicates that they have already passed 
the phase of active gas accretion. We can use the H$\alpha$ data for the three more distant sources
and the minimum observable $\dot{M}$ from magnetospheric accretion
models to infer an upper limit to the mass accretion rate.
Magnetospheric accretion models predict that the  minimum observable $\dot{M}$ 
depends on $R_\star^{(3/2)} \times M_\star^{(1/2)}$ and is about
10$^{-10}\,M_\sun/yr$  for a 1\,Myr old star with $M_\star$=0.5\,$M_\sun$
and $R_\star$=2\,$R_\sun$  (Muzerolle priv. comm.).
Given this dependence, the absence of H$\alpha$ in emission and the 
radii and masses of the stars, we can conclude that residual accretion  
may persist in these disks only at low rates $\lesssim 10^{-10}\,M_\sun/yr$.   
Accretion rates $\lesssim
10^{-10}$\,$M_\sun/yr$ translate into  $\Sigma_{\rm 1AU}\lesssim 1.4\,g/cm^2$ if
the disk is accreting with $\alpha=0.01$  and $T_{\rm 1AU}$ is 100\,K
(eq.~\ref{eq_accretion}). 
This surface density corresponds to a gas mass upper
limit of 0.001\,$M_{\rm J}$ from the magnetospheric radius
(3\,$R_\sun$=0.01\,AU) out to 1\,AU  when adopting our fiducial disk surface density of $\Sigma \propto r^{-1}$.
Although  $\alpha=0.01$  and $T_{\rm 1AU}$ are extremely uncertain (and that's why upper limits from {\it Spitzer} data 
alone are valuable), it seems very unlikely
that even the three more distant systems have large amounts of gas within 1\,AU. 
   Surface density limits at 0.3\,AU from the accretion are $\lesssim 5\,g/cm^2$, comparable to
{\it Spitzer} limits for nearby sources but much more stringent than what could be set from infrared data alone on 
these three more distant sources.

 In summary, we can exclude gaseous reservoirs of 0.04\,M$_{\rm J}$  within a
few AU of the disk inner radius for disk radii from 1\,AU up to $\sim$40\,AU. 
{\it Spitzer} data alone are sensitive to small amounts of gas from about
0.3\,AU to 1\,AU for the majority of the sources. For the three more distant
systems, accretion rate indicators set similarly stringent upper limits in the
inner disk regions ($<$1\,AU). For disks with inner holes larger than 40\,AU,
mid--infrared  transitions are not sensitive tracers of surface density because
the gas is too cold. However, CO millimeter non--detections can exclude the
presence of a large reservoir  of cold gas at $\ge$20\,AU.  In conclusion,
there is no indication that the targeted disks have enough gas  to form
Jupiter-- or  Saturn--like planet(s).  The present data cannot exclude the
possibility that some or all of these systems have  already formed such
planets.

What can we learn about the gas dissipation timescale  based on the above
discussion?   Our sample  includes sources that are beyond the phase of active
gas accretion. In addition, their small infrared excesses  suggest that they
have already dissipated most of their primordial dust and/or  agglomerated it
into larger particles. Our results show that by this time most of the gas
has also been dispersed perhaps by photoevaporation  (e.g.
\citealt{2001MNRAS.328..485C}), and/or accreted to form gas giant planets. 
These conclusions are consistent with a rapid gas dissipation timescale   
that leaves little trace of gas in optically thin dust disks.  More
statistics on sources  younger than $10\,$Myr in age and also on sources with
optically thick dust disks are necessary to observationally constrain the gas
lifetime and explore any link between gas and dust dissipation.

\begin{figure*}
 \resizebox{\textwidth}{!}{\includegraphics{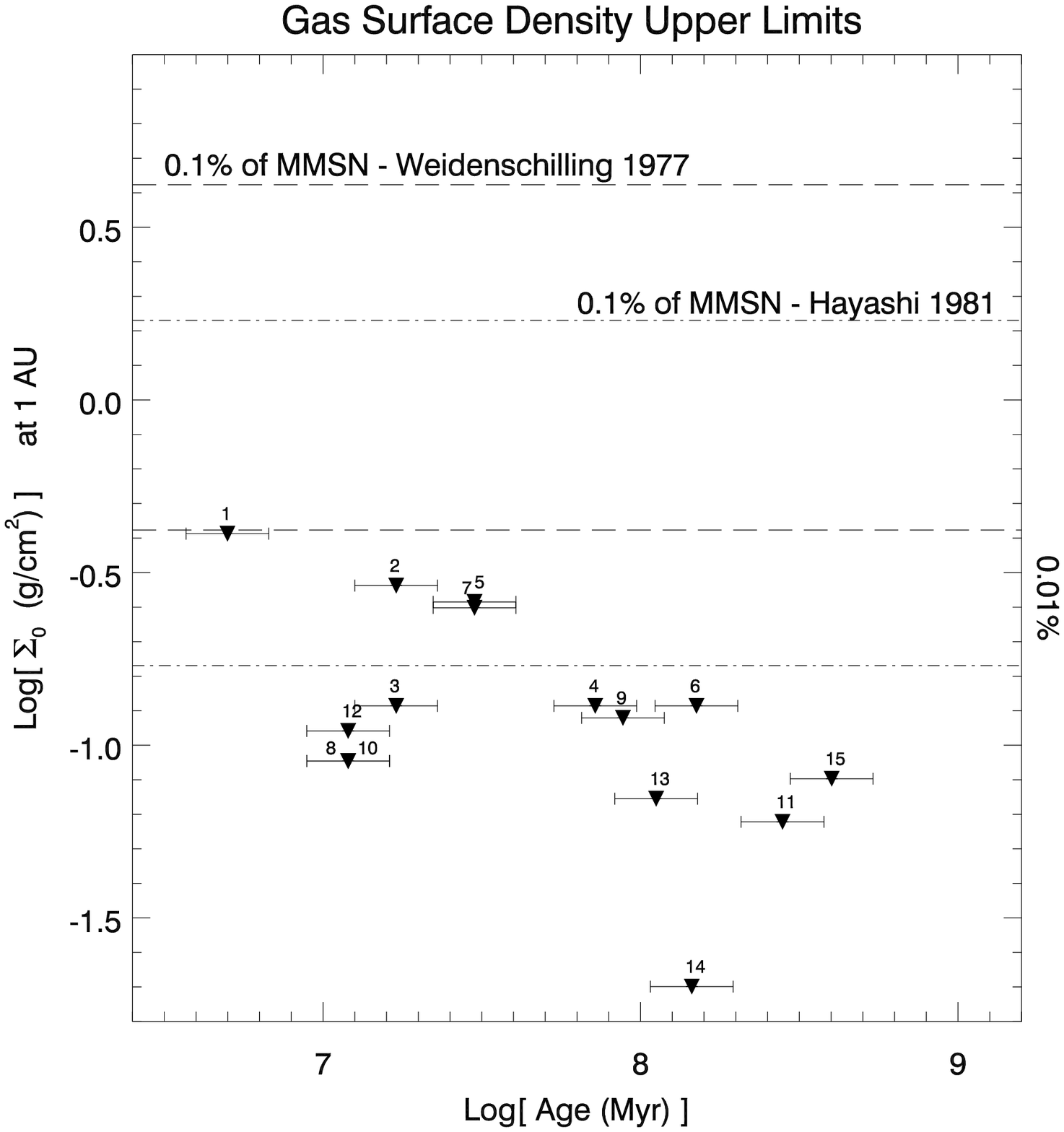}}
\caption{Gas surface density upper limits  versus age compared to the surface density of the MMSN model from \citet{1977Ap&SS..51..153W} (dashed lines)
and from \citet{1981PThPS..70...35H} (dash--dotted lines). 
Surface density upper  limits are from the   \si{} line for a disk with $R_{\rm in}$=1\,AU 
(see Table~\ref{gas_mod}), ages and errorbars are from Table~\ref{stars}.
Our surface density upper limits at 1\,AU are lower than 0.01\% of the MMSN value from 
\citet{1977Ap&SS..51..153W} for all sources. Note that the anti-correlation between the density upper limits at the inner radius and age does not depend on our model assumptions but rather on the properties of our sample. 
Because young sources are more distant than old ones, 
our line luminosity upper limits are less stringent for young disks 
(see Table~\ref{stars} and Fig.~\ref{figLineup}).}
\label{MMSN}
\end{figure*}

Short gas dissipation timescales are not only relevant to the formation of giant
planets but also to the later stages of terrestrial planet formation. These final
stages of growth may have involved tens to hundreds of planetary embryos over a
timescale of 10 to 100\,Myr \citep{2004ARA&A..42..441C}.  Numerical simulations find
that terrestrial planets grow to Earth-size in timescales  of tens of million of
years,  in agreement with isotopic constraints \citep{2002Natur.418..952K}. A problem
in the simulations of these later stages is that secular perturbations by (the
presumably already formed) Jupiter and Saturn and by neighboring embryos excite
terrestrial planet eccentricities\footnote{Collisions with large impactors (up to the
mass of Mars) occur frequently at average times of $\sim$30\,Myr and random
orientations but seem to cause smaller changes in the eccentricities than secular
perturbations (see e.g. \citealt{2001Icar..152..205C})}.  Thus, some damping mechanisms
seem to be required to circularize the final orbits of terrestrial planets. One
possibility is through tidal interactions with a remnant gas disk. 
\citet{2002Icar..157...43K} suggest that small amounts of gas, between 0.1-- 0.01\% of
the minimum mass solar nebula (MMSN,
\citealt{1977Ap&SS..51..153W,1981PThPS..70...35H}), can reduce the eccentricities to
values as low as those of Earth and Venus  on a timescale of about 10\,Myr if secular perturbations
from Jupiter and Saturn do not have a significant effect upon the evolution
\citep{2004Icar..167..231K}.  Alternatively, dynamical friction associated with remnant
swarms of planetesimals can damp the eccentricities over longer timescales even in
the presence of perturbations from the giant planets
\citep{2006LPI....37.2347O,2006....Raymond}. Thus our low gas surface density upper
limits at 1\,AU  (see Fig.~\ref{MMSN}) suggest that circularization of terrestrial
planets  if it occurs beyond 10\,Myr may have happened primarily through the latter
mechanism, although certainly gas could have played a role during the earlier stages of
terrestrial planet formation.

Finally, the gas limits in the 10-40\,AU region,  the region mainly
traced by our millimeter CO observations, may also be relevant
to the formation of outer gas-poor giant planets such as
Uranus and Neptune.
Because of the dynamical interactions between embryos and the gravitational 
effects of Jupiter and Saturn, very little accretion of solids is expected to have occurred 
at the location of Uranus and Neptune in the absence of mechanisms to increase capture cross 
sections of embryos \citep{2001Icar..153..224L}. One way to do this is through gas drag. 
However, this mechanism would require many Jupiter masses of gas 
(e.g. \citealt{2004ARA&A..42..549G}) that are not observed in any of our targets 
(see Tables~\ref{gas_mod} and \ref{gas_mod_co}). 
The CO data for eight of our sources (with ages between 12 and few hundred Myr) indicate 
that less than 2 Earth masses of gas are present between 10 and 40\,AU in our disks. 
These values are smaller than the gas mass of Uranus and Neptune as inferred by models of 
their interiors \citep{1999Sci...286...72G}. Thus  if Uranus and Neptune formed at their current radii, 
either gas persisted much longer in the 
solar nebula than was the case for our target stars, or they formed relatively 
quickly, in much less than 100\,Myr.

\section{Summary and Conclusions}\label{sect:summ}
We analyzed  infrared and millimeter spectra for 15 of our FEPS
sources selected for investigating the gas dispersal timescale in
disks around solar-type stars. Our targets span a wide age range  
from 5 to few hundred Myr (with 50\% of the sample younger
than $\sim$30\,Myr) 
and most are surrounded by optically thin dust
disks. We did not detect gas lines in the {\it Spitzer} IRS modules
nor in the millimeter with the SMT.
We estimated upper limits to the gas mass using simple approximations and also
more sophisticated gas and dust models. 
In agreement with  our previous modeling of HD~105 (H05), we find that 
optically thin dust disks have too little dust surface density
to appreciably heat the gas. Therefore, X--ray and UV flux from the central 
star become the main heating mechanisms for the gas. 
 We show that gas line upper limits from {\it Spitzer} provide
sensitive limits to the gas surface density at the disk inner radius. Millimeter
CO data are complementary by setting limits on the gas mass in the outer disk ($\ge$15\,AU).
We have also discussed
the robustness of our results by varying the main uncertain input parameters 
to the disk models. 
Future work will include a study weighted towards disks younger than 10\,Myr and
tests of the gas models on sources with detected gas lines probing different disk regions.
Our main conclusions can be summarized as follows:
\begin{enumerate}

\item Simple estimates of gas mass upper limits from mid--infrared H$_2$ lines indicate that even the five youngest disks (5-20\,Myr) in our sample have less than 0.6\,$M_{\rm J}$ of gas at 100\,K. 
Detailed gas models of the infrared and millimeter upper limits combined with the absence of accretion
 signatures allow us to conclude that none of the targeted disks have enough gas to
 form Jupiter-- or Saturn--like planets at the present time.

\item Our results do not support the presence of "large quantities of gas in
debris disks" as proposed by \citet{2001Natur.409...60T}. On the contrary, we
have evidence that systems with small dust excess in the 10--30\,Myr age range
do not have large amounts of remnant gas.

\item We estimate gas surface density upper limits at 1\,AU smaller than 0.01\% of
the MMSN model for most of the sources, eight of which have ages between
5--30\,Myr.  If the circularization of terrestrial planets occurs in this age range, then gas surface densities appear to be so low that interactions with planetesimals
seem to be the only viable mechanism to circularize their orbits.

\item The gas limits from CO data in the 10-40\,AU region are less than a few
$M_\oplus$. These values are far too low for gas drag to enhance the gravitational 
cross section of embryos and speed up the in-situ formation of Uranus and Neptune. 
In addition, if these systems are analogs of the Solar System, our results indicate relatively short timescales ($<<$\,100\,Myr) for the formation of  Uranus-- and Neptune--like planets.

\end{enumerate}

\acknowledgments
It is a pleasure to thank all members of the FEPS team for their 
contributions to the project and to this study.
IP wishes to thank D. Watson for suggestions in the data reduction of 
the IRS high-resolution spectra and J. Muzerolle for helpful discussions 
on the mass accretion rate in young circumstellar disks.
We would also like to thank the anonymous referee for a very careful and helpful review.
This work is based on observations made with the Spitzer Space Telescope, 
which is operated by the Jet Propulsion Laboratory, California Institute 
of Technology under NASA contract 1407. 
Support for this work was provided by NASA through the FEPS Legacy award 
issued by JPL/Caltech.


Facilities: \facility{Spitzer Space Telescope}.

\bibliography{lit}

\clearpage

\end{document}